\documentclass[useAMS,usenatbib,usegraphicx]{mn2e}
    \newcommand{\Teff}{\mbox{$T_{{\rm eff}}$}}  % Teff

        \def\smallskip{\vskip 4pt}
        
\def\apjs{ApJS}%
\begin{document}
\title[The {\it GALEX} UV emission in shell galaxies]
{The {\it GALEX} UV emission in shell galaxies: tracing galaxy ``rejuvenation" episodes}
\author[R. Rampazzo et al.]{R. Rampazzo$^{1}$\thanks{E-mail:
roberto.rampazzo@oapd.inaf.it}, A. Marino$^{1}$, R. Tantalo$^{2}$, D. Bettoni$^{1}$, L.M. Buson$^{1}$, C. Chiosi$^{2}$,
\newauthor  G. Galletta$^{2}$, R. Gr\" utzbauch$^{3}$ and R.M. Rich$^{4}$  \\
$^{1}$INAF-Osservatorio Astronomico di Padova, Vicolo dell'Osservatorio 5, 35122 Padova (Italy)\\
$^{2}$Dipartimento di Astronomia Universit\'a di Padova, Vicolo dell-Osservatorio 2, 35122 Padova (Italy)\\
$^{3}$Institut f\"ur Astronomie, Universit\"at Wien, T\"urkenschanzstra{\ss}e 17, A-1180 Wien, Austria\\
$^{4}$Physics \& Astron. Dept., UCLA, Box 951562, 405 Hilgard Ave., Los Angeles CA 90025-1562, USA
}

\date{Accepted. Received; in original form}
\pagerange{\pageref{1}--\pageref{21}} \pubyear{2007}

\maketitle

\label{firstpage}

\begin{abstract}

We present the {\it Galaxy Evolution Explorer (GALEX)} far (FUV) and
near (NUV) ultraviolet imaging of three nearby shell galaxies,
namely NGC~2865, NGC~5018, and NGC~7135 located in low density
environments.

The system of shells and fine structures visible in the optical is
detected in the NUV image of NGC~2865  and  in both NUV and FUV
images of NGC~7135. The NUV image of NGC~5018  does not present
shell structures. We detect absorption features in the nuclear
region of all three galaxies. NGC~2865 has a nearly flat colour
profile with (FUV-NUV)$\approx$2 throughout the whole galaxy.
NGC~7135 is blue in the center (FUV-NUV)$\approx$0 and as red as
(FUV-NUV)$\approx$1.5 in the outskirts, including the faint
shell-like feature.

The three shell galaxies are members of poor groups of galaxies. We
compare {\it GALEX} NUV observations with available H\,{\sevensize I}
large scale measurements, and determine the UV magnitudes of likely
companions.  Most of the known (and possible) companions are gas rich
late-type galaxies, suggesting that our shell galaxies inhabit the
ideal environment for hosting {\it rejuvenating} episodes driven by
accretion events.

We investigate the ability of the {\it nuclear} {\it GALEX}
(FUV-NUV) colour to provide information about {\it rejuvenation}
phenomena in the stellar populations of the shell galaxies. To this
aim, we derive from theory the relationship between the Mg2,
H$\beta$, H$\gamma$A, H$\delta$A Lick line-strength indices and the
(FUV-NUV) colour. We extend the study to a sample of early-type
galaxies in low density environments which includes shell galaxies
and/or galaxies with emission lines in their optical spectra
\citep{Annibali07}. In the index vs. (FUV-NUV) colour diagrams, most
of the galaxies are well explained by passively evolving Single
Stellar Populations. On the average, ages and metallicities
of the galaxies in our sample estimated from optical line-strength
indices are consistent with those inferred from the (FUV-NUV)
colour. We note that the {\it GALEX} (FUV-NUV) and (UV-V) colours
have different response to age and metallicity. In general, all the
colours but for (FUV-NUV) and (FUV-V), become nearly age insensitive
when 1-2 Gyr have elapsed from the last star forming event. Finally,
considering composite stellar population models with a recent burst
of star formation,  we suggest that the position of the NGC~7135 and
NGC~2865 nuclei in the (FUV-NUV)-H$\beta$ plane could be explained
in term of a recent rejuvenation episode.
\end{abstract}


\section{Introduction}

Shell galaxies represent the ideal class of objects to investigate
galaxy evolution in the field for several reasons. Two different
scenarios for their origin emerge from the rich harvest of simulations
performed since their discovery in the early 80's: either a weak
interaction between galaxies \citep{Thomson90,Thomson91} or
merging/accretion events between galaxies of different masses (mass
ratios typically 1/10 - 1/100; \cite{Dupraz86,Hernquist87a,Hernquist87b}).

Weak interactions can form long lasting azimuthally distributed
shells through the interference of density waves produced in a thick
disc population of dynamically cold stars.  However, this requires a
cold thick disc, not found in ellipticals. In merging models, shells
are density waves formed by  infall of stars from a companion
during a minor merger. Major merger can also  produce shells
\citep{Barnes92,Hernquist92,Hernquist95}.

Both scenarios qualitatively reproduce basic characteristics, such
as spatial distribution, frequency and shape of observed shell
systems \citep[see e.g.][and references therein]{Wilkinson00}. Weak
interaction models add also clear information about the environment:
shells could not be maintained within clusters, as the continuous
``harassment'' among galaxies would destroy them.

Shell galaxies indeed avoid the cluster environment and are found in
the field with a high frequency ($\approx$ 16.5\% of the early--type
galaxy population, \citet{Malin83,Schweizer92,Reduzzi96,Colbert01}. This suggests that
interaction/accretion/merging events have played a significant role in
the formation/evolution of the early-type class as a whole.

Using Fabry-Perot observations, \citet{Rampazzo03} analyzed the warm
gas kinematics in a few shell galaxies. They found that gas and stars
appear to be decoupled in most cases. This evidence suggests an
external acquisition of the gas, as predicted by merging models.  At
the same time, a set of observations showing a clear association
between cold (HI/CO) gas and stars challenge present merging models
which do not predict it unless cold gas behaves differently from the
ionized gas \citep{Schiminovich94,Schiminovich95,Charmandaris00,Balcells01}.
Studying the distribution of shell galaxies in the H$\beta$ vs [MgFe]
line-strength index plane, \citet{Longhetti00} show that these systems
span a wide range of ages, indicating that among them recent and old
interaction/acquisition events are equaly probable.  If shells are
formed at the same time at which the rejuvenating event took place,
shells ought to be long--lasting phenomena. So, whereas current
studies in literature considers the shell structure surrounding
early-type galaxies as a {\it bona fide} indicator of past
accretion/merging events, their formation age and secular evolution
are far from being firmly established.

In the above framework we discuss {\it GALEX} observations of three
shell galaxies taken from the \citet{Malin83} compilation, namely
NGC~2865, NGC~5018 and NGC~7135. The galaxies are located in very
low density environments and belong to HI rich associations.
NGC 2865 and NGC 7135 are part of our {\it GALEX} proposal
(GI04-0030-0059). The NGC5018 data are taken from the {\it GALEX}
archive. The line-strength indices analysis performed by
\citet{Longhetti00} and \citet{Annibali07} suggests that the above
shell galaxies had a recent burst of star formation.  The present
sample is then not representative of the entire class of shell
galaxies \citep[see e.g. ][]{Longhetti00}. Through this pilot study
we aim to verify whether these galaxies host ongoing star formation
and how it distributes across the galaxy. 

The plan of the paper is as follows. Section~2 describes the relevant
properties of our galaxy sample that could be gathered from
literature. Section~3 presents the {\it GALEX} observations and the
data reduction. Results are presented in Section~4 and discussed in
Section~5.

%-------------------------------------Table 1--------------------------
\begin{table*}
\caption{ Data for the three shell galaxies observed in this study }
\begin{tabular}{llllc}
\hline\hline
                                  &  NGC~2865    & NGC~5018      & NGC~7135   & Ref. \\
\hline
Morphol. Type                     &     E3-4     &   E3          & SA0-pec    & [1] \\
Hel. Sys. Vel. [km~s$^{-1}$]      &  2627$\pm$3  &   2794$\pm$15 &2640$\pm$21 &  [1] \\
Adopted distance [Mpc]            & 35.7         & 40.9          & 34.7       &  [2]\\
$\rho_{(x,y,z)}$ [gal Mpc$^{-3}$] & 0.11         & 0.29          & 0.32       & [2] \\
Environment                       & Antlia-Hydra clouds & Virgo Southern extension & Pisces-Austrinus Spur & [2] \\
                                                  &                         &                      & &\\
{\bf Apparent magnitude}          &              &               &            &\\
{\bf and colours}:                &              &               &            &\\
B$_T$                             & 12.57$\pm$0.14  &  11.69$\pm$0.13 & 12.79$\pm$0.15& [1] \\
$\langle$(B-V)$_T \rangle$       &   0.91   &  0.92   & 0.99 &      [1] \\
$\langle$(U-B)$_T \rangle$       &   0.41   &  0.48   & 0.42 &      [1] \\
(J-H)$_{2MASS}$                    &  0.65   &  0.63   & 0.66 &      [1]\\
(H-K)$_{2MASS}$                   &  0.25   &  0.30   & 0.22  &     [1] \\
                                                 &                            &              &  & \\
{\bf Galaxy structure}:                  &                             &             &  & \\
Effective Surf. Bright. $\mu_e$(B) &  20.20$\pm$0.27  & 20.35$\pm$0.25
&22.78$\pm$0.60&[3]\\
Diam. Eff. Apert., A$_e$ [\arcsec] &  25.0  & 45.5 &   62.8 &[3] \\
Avearage ellipticity      & 0.17 & 0.32 & 0.28 & [3]\\
P.A. [deg]      & 154.6   & 99.9 & 44.9 & [3]\\
                                                &                          &                      &   &     \\
{\bf Kinematical parameters}        &                         &                       &    & \\
Vel.disp. $\sigma_0$ [km~s$^{-1}$]  stars   &  180$\pm$24& 240 &202$\pm$21&[3] \\
Max. rotation V$_{max}$  [km~s$^{-1}$]      & 234$\pm$30&75$\pm15$& 78&[3] \\
                          &                          &                    &  & \\
\hline
\end{tabular}
\label{table1}

\medskip
References: [1] {\tt NED http://nedwww.ipac.caltech.edu/} ; [2] \citet{Tully88} (H$_0$=75 km~s$^{-1}$~Mpc$^{-1}$);  [3] {\tt HYPERLEDA http://leda.univ-lyon1.fr/}.
\end{table*}
%--------------------------------end table 1---------------------------

%----------------------------------------------Table 2 -------------------------------------------
\begin{table*}
\caption{Journal of the {\it GALEX} observations}
%\begin{tabular}{lcclllll}
\begin{tabular}{lcclll}
 \hline
 \multicolumn{1}{l}{Name}&
\multicolumn{1}{c}{NUV}  &
\multicolumn{1}{c}{FUV}  &
\multicolumn{1}{l}{observing} &
\multicolumn{1}{l}{P.I.} \\
%\multicolumn{1}{c}{$\mu_{sky}^{NUV}$}&
%\multicolumn{1}{c}{$\mu_{sky}^{FUV}$}\\
\multicolumn{1}{c}{}&
\multicolumn{1}{c}{ exposure [sec]} &
\multicolumn{1}{c}{ exposure [sec]} &
\multicolumn{1}{c}{date}&
\multicolumn{1}{c}{} \\
%\multicolumn{1}{l}{mag~arcsec$^{-2}$}&
%\multicolumn{1}{l}{mag~arcsec$^{-2}$}\\
\hline
%NGC~2865  &16269       & 2567       &2004-12-22       & D. Bettoni &  26.26 & 26.90 \\
%NGC~5018  & 1238 & 118        &$^{*}$2005-05-14 & $^{*}$ J. van Gorkom & 26.46&\\
%NGC~7135  &1693        &1693        & 2004-10-15      & D. Bettoni& 26.65 & 26.89 \\
NGC~2865  &16269       & 2567       &2004-12-22       & D. Bettoni \\
NGC~5018  & 1238$^{*}$ & 118    &2005-05-14$^{*}$ & J. van Gorkom$^{*}$  \\
NGC~7135  &1693        &1693        & 2004-10-15      & D. Bettoni \\
\hline
\end{tabular}
\label{table2}

\medskip{Note: The observing date and the P.I. labeled with an asterisk refer to the NUV observation of NGC~5018.}
\end{table*}
%---------------------------------------------------end Table 2-----------------------------------------------------

%--------------------------Figure 1----------------------------------
\begin{figure*}
{
{\includegraphics[width=14.4cm]{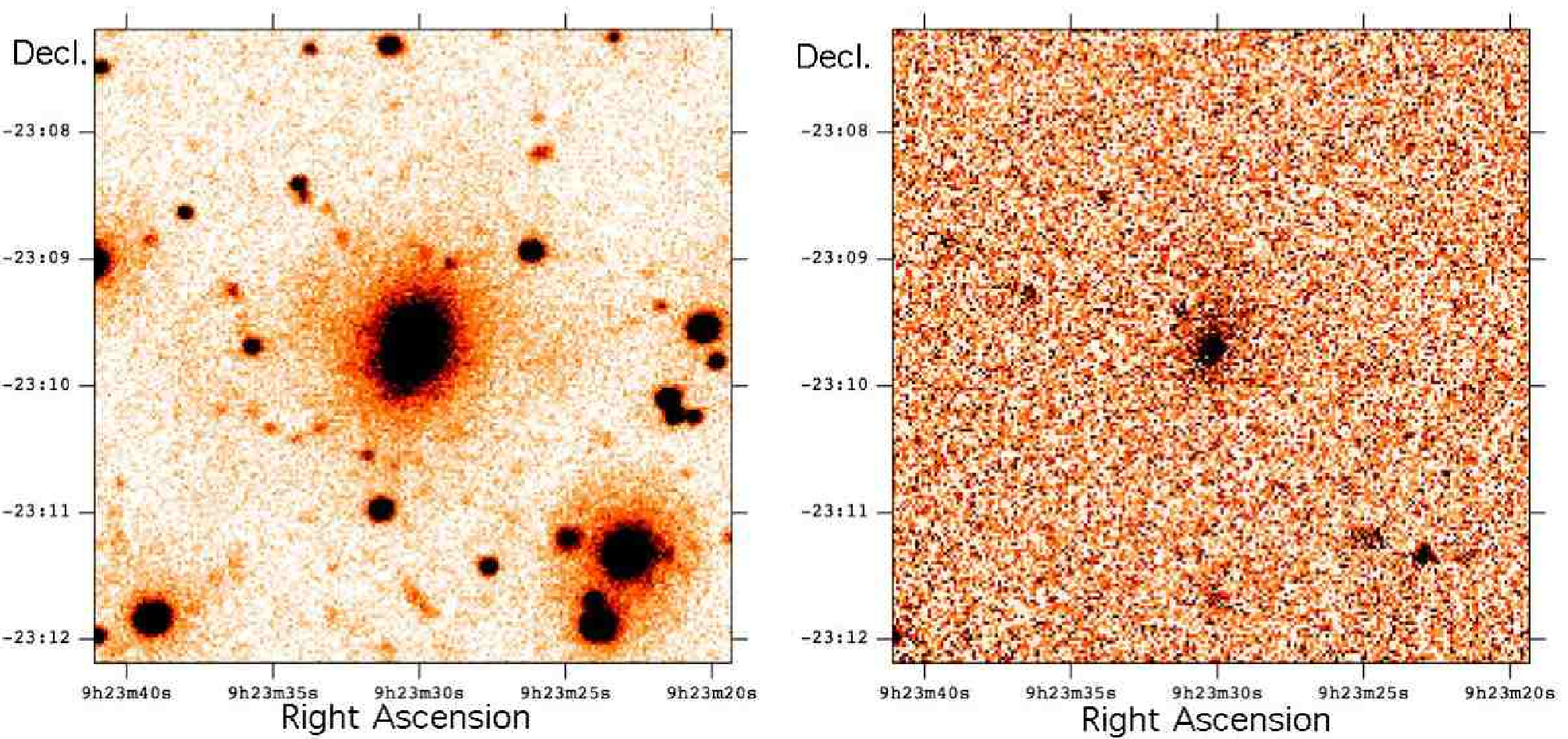}
\includegraphics[width=14.4cm]{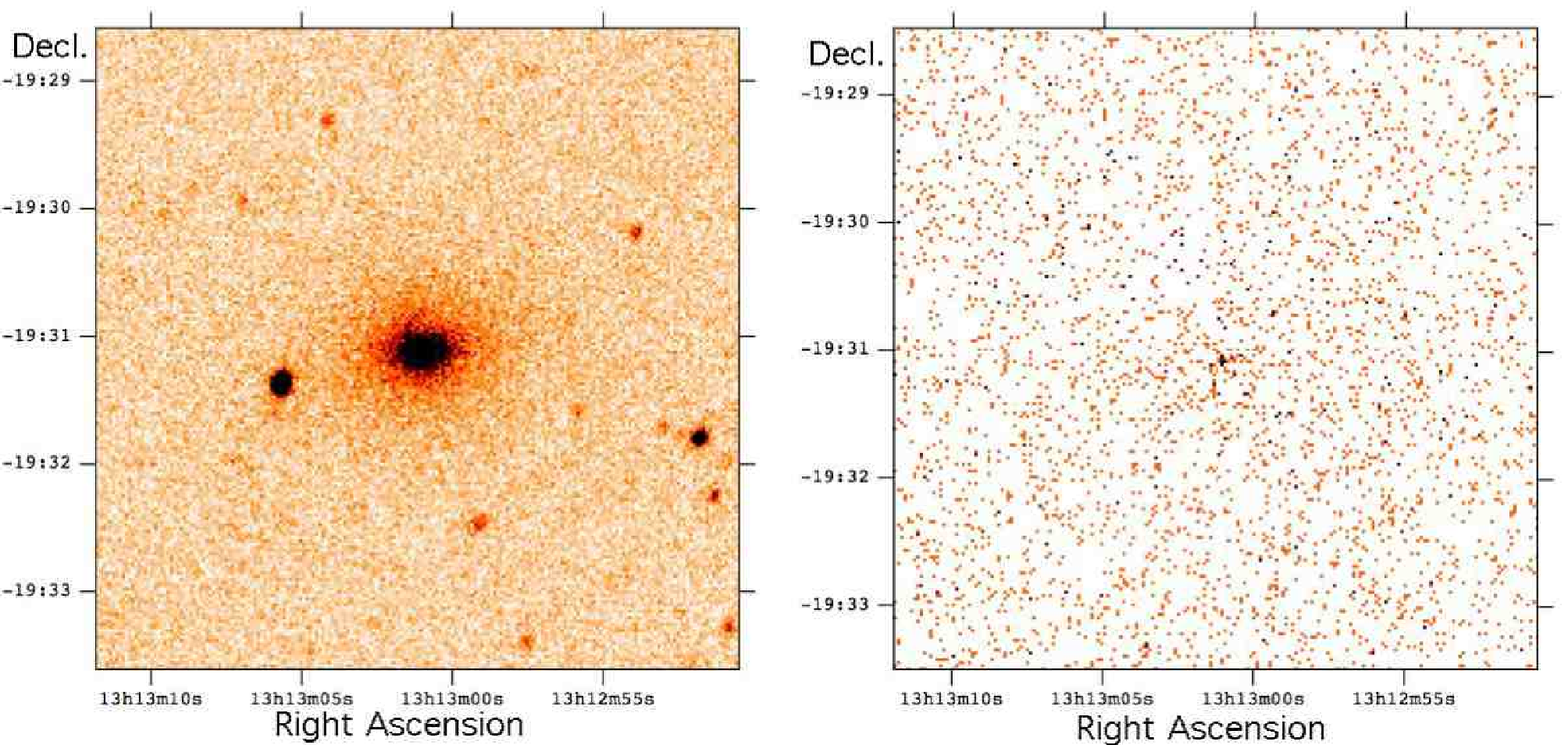}
\includegraphics[width=14.4cm]{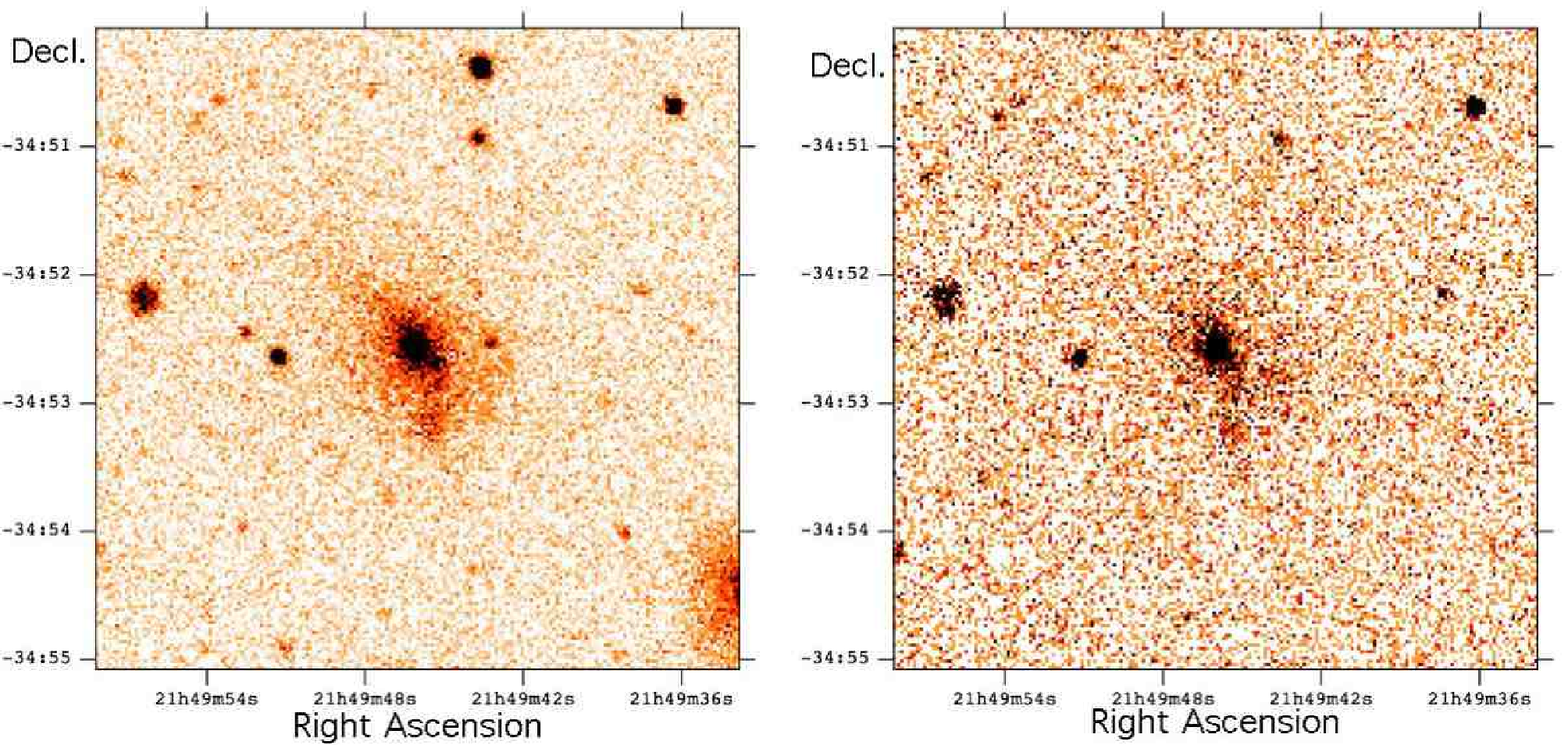}}
}
\caption{NGC 2865 (top panels),  NGC 5018 (mid panels) and NGC 7135 (bottom panels)
full resolution, 5\arcmin $\times$ 5\arcmin\ wide {\it GALEX} NUV (left panels) and FUV
(right panels), background subtracted images  in counts~px$^{-1}$~s$^{-1}$.
\label{fig1}}
\end{figure*}
%--------------------------end Figure1-------------------------------------

\section{The sample}

We collect in Table~\ref{table1} the main characteristics of the
galaxies under examination.  As may be deduced from the local galaxy
density, $\rho_(x,y,z)$ [gal~Mpc$^{-3}$], the galaxies are located in
very low density regions of complex galaxy associations
\citep{Tully88}.

\noindent{\bf NGC 2865}. According to  \citet {Jorgensen92} the
galaxy is a {\it bona fide} elliptical with a surface brightness
profile consistent with a pure r$^{1/4}$ law. At the same time,
NGC~2865 is genuinely peculiar: in deep images the galaxy displays a
significantly disturbed morphology. The isophote fit reveals boxy
isophotes within 40\arcsec and disky isophotes at larger radii
\citep{Reid94}. The galaxy displays a chaotic system of about 7
shells out to 2.5\arcmin\ \citep{Fort86}. A westward protrusion,
identified as a stellar jet or a polar ring \citep{Whitmore90},
extends 1\arcmin\ from its nucleus. \citet{Fort86} estimate that
11-22\% of the total luminosity of NGC~2865 is contained in the shells
and in the above described protrusion.  Further out, in deep images, a
faint loop is visible north west and a plume or tail-like extension
can be seen to the south east.  Nevertheless, \citet{Schiminovich95}
HI observations show that NGC~2865 has no companions of similar
luminosity, while two gas-rich dwarfs are seen nearby. The above
authors suggest that the galaxy might be the recent (less than 7 Gyr)
product of a major disk-disk merger, although the association between
gas and stellar fine structure, with gas displaced outward from the
stars in projected position, implies gas motions not predicted by any
of the current merger scenarios. Using nuclear line-strength indices
\citet{Longhetti99} suggest that NGC 2865 is among those shell
galaxies which are candidate for having a very young stellar
component in their spectra: their analysis of both [CaII] and
H$\beta$ line-strength indices provide a luminosity weighted
stellar age lower than 2 Gyr.

\noindent{\bf NGC~5018}. The galaxy is peculiar in two main
respects. First \citet{Schweizer90} classified NGC 5018 as one of his
prime candidates for a recent major merger, assigning to it a fine
structure $\Sigma$ parameter of 5.15. The $\Sigma$ parameter is an
empirical measure of the optical disturbance present in a galaxy since
its value is given by a combination of the optical strength of
ripples, the number of detected ripples, the number of jets, an
estimate of boxiness and the presence of X-structures.  The higher the
value of $\Sigma$ the higher the morphological disturbance and the
probability that the galaxy is {\it dynamically younger}.  Second, the
nuclear optical spectrum has the weakest Mg2 line-strength index
considering its velocity dispersion among the over 400~Es surveyed by
\citet{Davies87}. Also unusual is the lack in its UV IUE spectrum
\citep{Bertola93} of a prominent UV-upturn short ward of $\lambda$
2000~\AA, that is typical of old, metal-rich spheroids
\citep{Burstein98}. Such an anomalous result inspired some authors
to propose a kind of conspiracy, where young stars and heavy dust
absorption combine with other to dilute the underlying Mg2
line--strength index, so as to turn a flat, young-star-dominated UV
energy distribution into the observed NGC~5018 UV-weak spectrum
\citep{Carollo94}.  This scenario, however, is not consistent with the
average internal reddening as low as E(B-V)$\sim$0.02 measured by
\citet{Buson01}. The relatively young nature of this object was
later  confirmed by \citet{Leonardi00}, who proposed that NGC~5018
is an old, metal-rich elliptical which has undergone a major
accretion event at some recent stage of its evolution. In this
framework, the above authors interpret NGC~5018 as an object
presently dominated by an acquired stellar population not older than
2.8~Gyr. A slightly different view is  proposed by \citet{Buson04}
who suggested that in NGC~5018 the bulk of its stellar population
were {\it formed ex-novo} about 3~Gyr ago, just at the time of a
major merging event. Finally \citet{Tantalo04a} explored the
possibility that bulk population of stars in NGC~5018 is old and
that its unusual high value of $H_\beta$ simply reflects the
rejuvenation that occurred about 2 Gyr ago, most likely by a merger,
which however was limited to a minor percentage of the stellar mass.

\noindent{\bf NGC~7135}. \citet{Malin83} noticed that the galaxy
presents``a curious jet and a shell" and a prominent tail on the
opposite side of a sharp curved feature. \citet{Keel85} classified
the galaxy as  a merger galaxy with a LINER nucleus. Both the
stellar tail and the elongated asymmetric distribution of the
ionized gas, detected by \citet{Rampazzo03} suggest indeed a recent
acquisition event. In addition the inner stellar radial velocity, measured
by \citet{Longhetti98}, has a similar profile but shifted by about
200 km~s$^{-1}$  with respect to the barycentric velocity of the
ionized gas.

\citet{Annibali07} derive a luminosity weighted age of 2.2 Gyr for
NGC7135, which suggests the presence of a young burst of 
star formation (SF hereafter).
NGC~7135 is included among the few prototypical galaxies
that might have suffered from a nearly radial merger \citep{Aalto00}.
Once the effects caused by the stellar burst will be over, NGC~7135
is expected to closely resemble  the so-called Medusa galaxy
(NGC~4194) as suggested by \citet{Rampazzo03}.

In conclusion, there is evidence that all the three galaxies might
have undergone  recent (2-3~Gyr) {\it rejuvenation} episodes. The
correlation between the onset of star formation and the event that
triggered the formation of the shell structure could be further
investigated using  ultraviolet morphology and photometry (i.e. NUV
and FUV {\it GALEX} observations).

\section{Observations and data reduction}

The {\it GALEX} mission and instruments are fully described  in
\citet{Martin05} and \citet{Morrissey05}. The observing logs,
including exposure times, for each galaxy are provided in
Table~\ref{table2}. The spatial resolution of the images is
$\approx$4\farcs 5 and 6\farcs 0 FWHM in FUV (1350 -- 1750~\AA) and
NUV (1750 -- 2750~\AA) respectively, sampled with 1\farcs
5$\times$1\farcs 5 pixels.

Two galaxies, namely NGC~2865 and NGC~7135, have been imaged with
{\it GALEX} by means of dedicated  observations awarded to our team.
 The NUV and FUV data for NGC~5018 have been taken from the {\it
GALEX} archive: the FUV image is part of the All Sky Survey (AIS).
Full resolution images in the NUV and FUV bands are shown in
Figure~\ref{fig1}. The quality of the images varies significantly as
the exposure time  ranges from the NUV image of NGC~2865, the
results of the co-adding of several images for a total exposure time
of $\approx$ 16ks, to the FUV image of NGC~5018 with  118 $s$.

Our first analysis addresses a morphological study of the galaxy
structures. In order to enhance the S/N in the galaxy outskirts and to
bring into evidence possible faint structures in the UV emission we
used both the adaptive kernels smoothing algorithm available in {\tt
IRAF}\footnote{{\tt IRAF} is distributed by the National Optical
Astronomy Observatories, which are operated by the Association of
Universities for Research in Astronomy, Inc., under cooperative
agreement with the National Science Foundation.}  and {\tt ASMOOTH}
\citep{Ebeling06}. For each individual pixel the algorithm increases
the smoothing scale until the S/N within the kernel reaches a preset
value. Thus, noise is suppressed very efficiently, while at the same
time the real structure, i.e. signal that is locally significant at
the selected S/N level, is preserved at all scales. In particular,
extended features in noise-dominated regions are visually enhanced
(see Figures~\ref{fig2}, \ref{fig3} and \ref{fig4}).

Surface photometry (Figure~\ref{fig6}) was carried out, on the
background subtracted images, with the {\tt ELLIPSE} fitting routine
in the {\tt STSDAS} package of {\tt IRAF} and with the {\tt GALFIT}
package \citep{Peng02}. The necessary photometric zero points were
taken from \citet{Morrissey05}.  AB magnitudes, photometric errors
and FUV-NUV colours were determined from the original un-smoothed
data. {\tt ELLIPSE} computes a Fourier expansion for each successive
isophote \citep{Jedrzejewski87}, resulting in the surface
photometric profiles. {\tt GALFIT} was used to perform a bulge-disk
decomposition -- if needed -- and to determine the parameters of the
Sersic profile fitted to the bulge of the galaxies. The Sersic
profile is a generalisation of the de Vaucouleur law with $\mu(r)
\sim r^{1/n}$, where $n$ is a free parameter \citep{Sersic68},
otherwise named the Sersic parameter. The profile is sensitive
to structural differences between different kinds of early-type
galaxies and thus provides a better fit to real galaxy profiles.
{\tt GALFIT} was also used to search for fine structures on the
HST-ACS F606W image of NGC~2865 obtained from the HST archive (see
Figure~\ref{fig2} central panel).

We finally analyzed the colour distribution across the galaxies. The
NUV and FUV images  have been scaled to the same spatial resolution
in order to avoid spurious colour gradients in the inner parts
(Figure~\ref{fig6}).

\section{Results}

\subsection{Galaxy morphology in NUV and FUV}

\noindent{\bf NGC~2865}

\begin{figure}
\includegraphics[height=6.2cm,width=6.9cm]{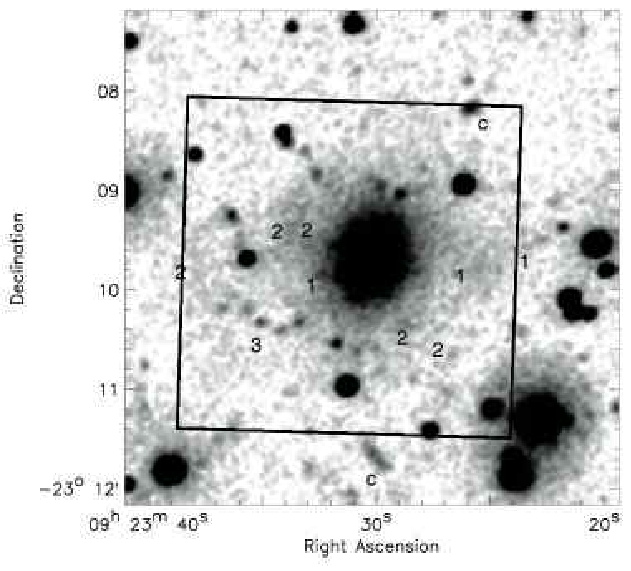}
\includegraphics[height=6.3cm,width=6.7cm]{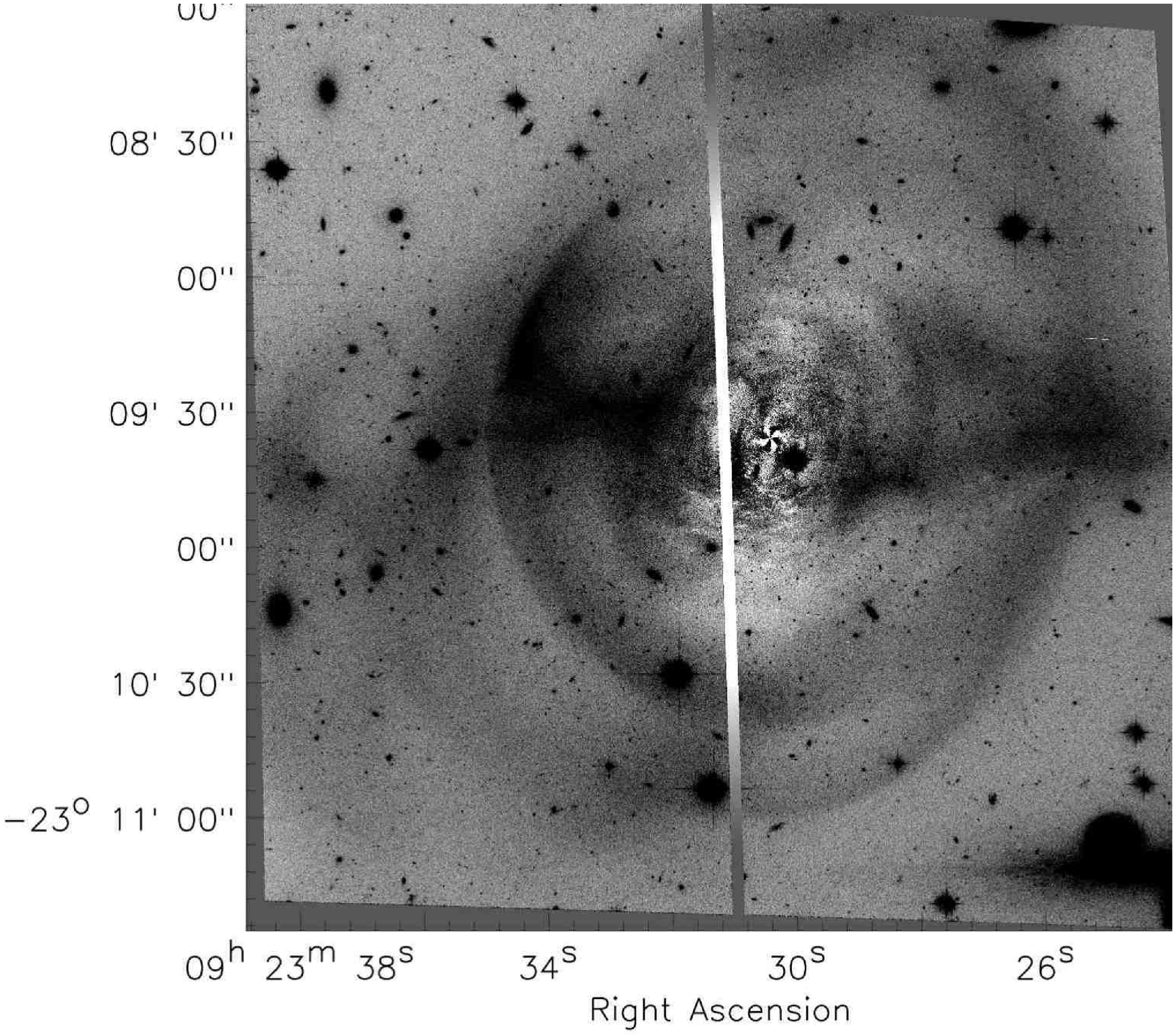}
\includegraphics[height=6.2cm,width=6.9cm]{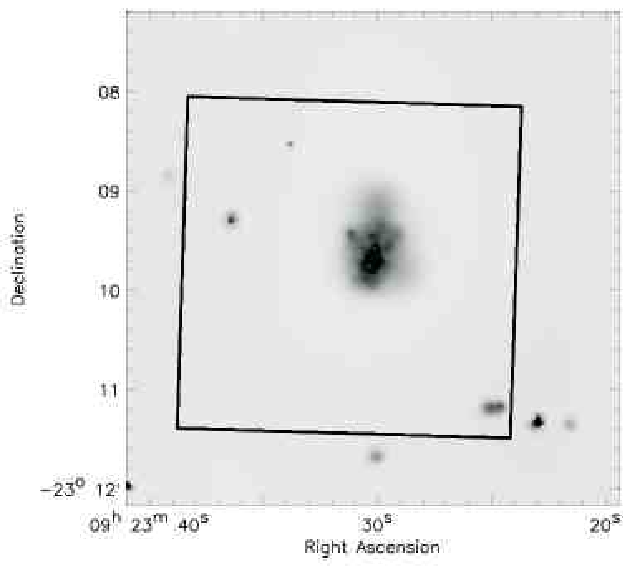}
\caption{NGC 2865. ({\it Top panel}) {\it GALEX} NUV image
(FOV= 5\arcmin$\times$5\arcmin).  The image has been smoothed with an
adaptive filter to enhance the outer fine structures.  We used for
comparison the HST-ACS F606W image of NGC~2865
(FOV=3.5\arcmin$\times$3.5\arcmin) ({\it mid panel}) where the main
body of the galaxy has been subtracted revealing a complex fine
structure and a central dust system.  On the NUV image, we labeled the
NE and NW protrusions ($1$) and the shell system ($2$) visible in the
HST-ACS field of view \citep[see also][see Figure~4 in the quoted
paper]{Fort86}. We further labeled with $3$ a chain of possibly
background galaxies barely visible in the optical band and with $c$
the galaxies indicated as ``companions" in \citet{Fort86}.  For the
faint FUV image {\it bottom panel}) we used the {\tt ASMOOTH}
(Adaptive Gaussian kernel) filter with $\tau_{min}=$3. The box in
the NUV and FUV images marks the HST-ACS FOV.
\label{fig2}}
\end{figure}

The extension of NUV emission (Figure~\ref{fig2} top panel) is
comparable with that of the optical image, whereas the FUV emission,
obtained with a much shorter exposure time (see Table~2), shows up
only in the central regions of the galaxy.

We attempted to enhance the fine structure in both NUV and FUV images
using adaptive filters. While the FUV image is dominated by the
presence of the dust visible in the optical image (see the central
panel of Figure~\ref{fig2}), several of the fine structures detected
in this latter are also discernible in the NUV image. Both
``protrusions'' in the West and East sides of the galaxy (evident in
the HST-ACS optical image) and some of the shells (labeled $1$ and $2$
respectively in the top panel) are detected. An external shell at
about 2\arcmin\ East of the nucleus is barely visible both in the NUV
and HST-ACS image.

The field is crowded with faint, possibly background galaxies. The two
galaxies, indicated as ``Companions'' in the sketch of fine structure
by \citet[][see their Figure 4]{Fort86} are detected in NUV and
labeled with $c$. The coordinates of the northern object, catalogued
as 2MASX J09232585-2308093, are 09 23 25.8 -23 08 09. The magnitude is
mag$_{NUV}$=22.37$\pm$0.08. The southern un-catalogued object is
located at 09 23 29.97 -23 11 41.9. The magnitude is
mag$_{NUV}$=21.09$\pm$0.04.

%--------------------------Figure 2----------------------------------
%--------------------------end Figure2-------------------------------------

%--------------------------Figure 3----------------------------------
\begin{figure}
\includegraphics[height=6.7cm,width=7.6cm]{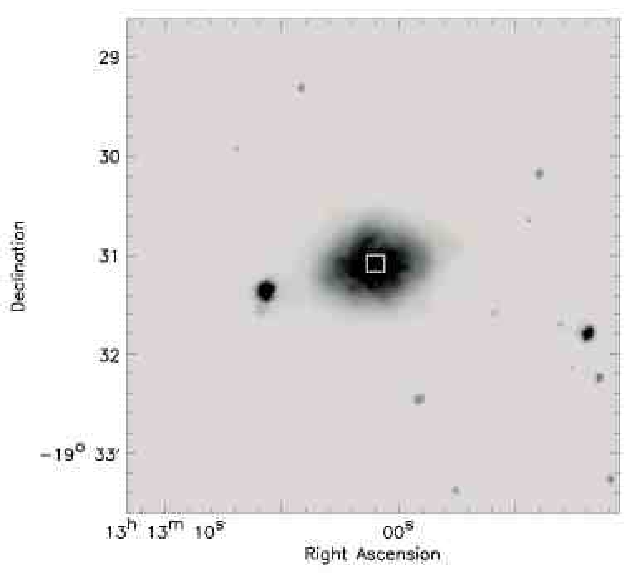}
\includegraphics[height=6.7cm,width=7.6cm]{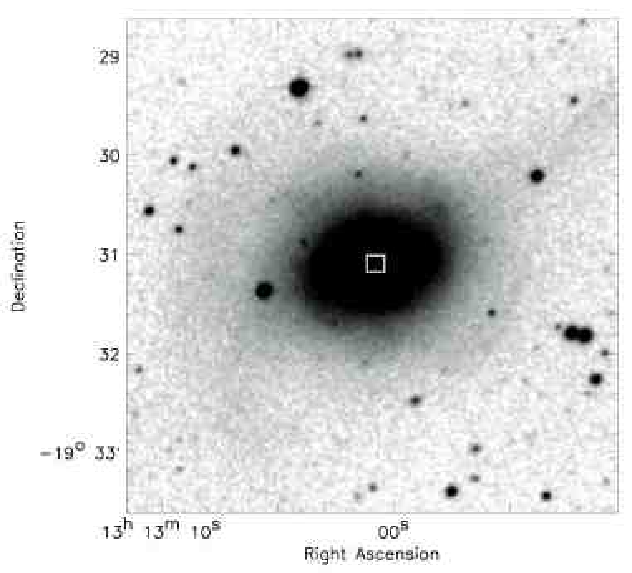}
\includegraphics[height=7.3cm,width=7.3cm]{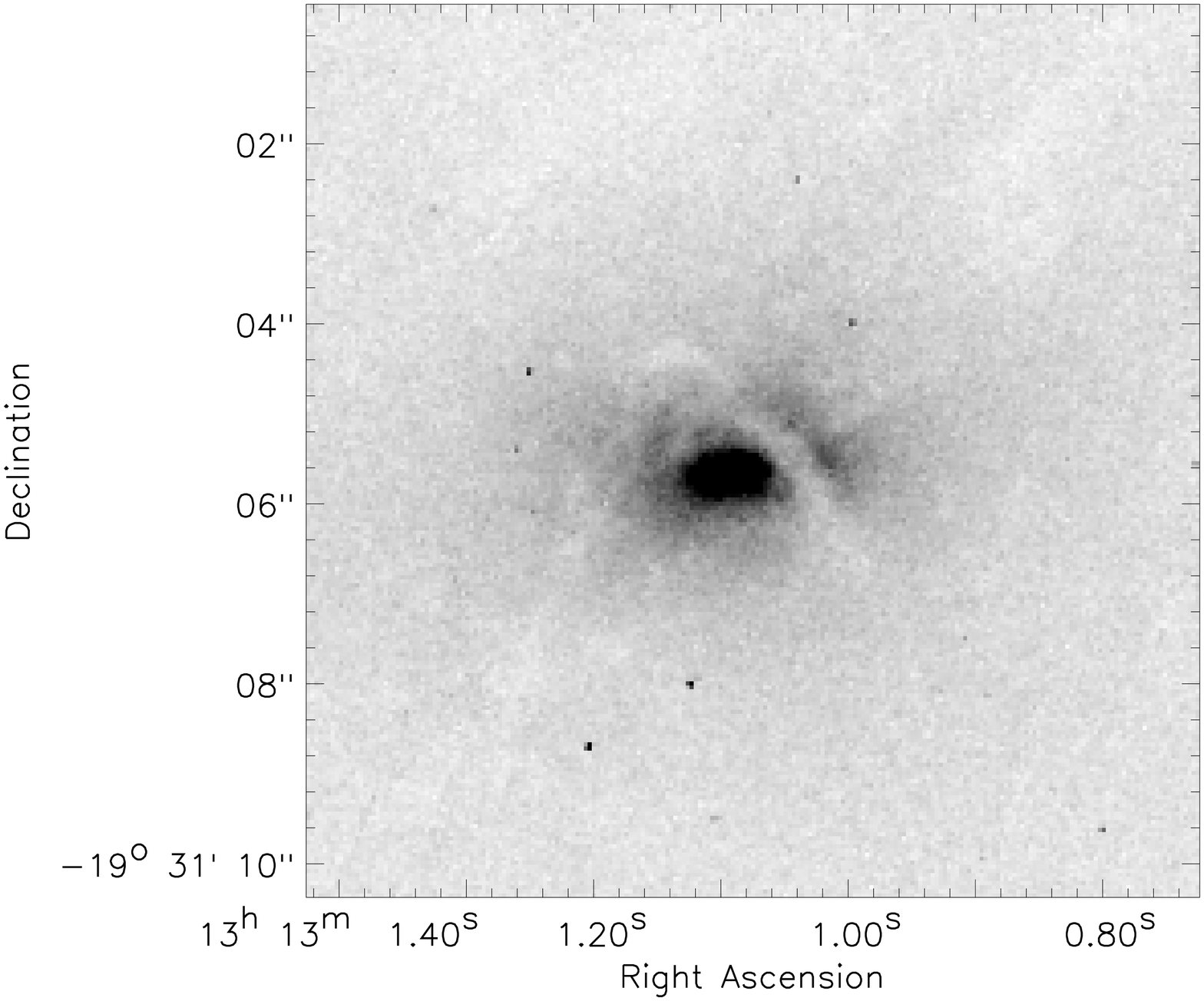}
\caption{NGC 5018. ({\it top  panel}) {\it GALEX} NUV (5\arcmin$\times$5\arcmin)
image after the application of {\tt ASMOOTH} filter with $\tau_{min}=$
7 .  ({\it mid panel}) UKSTU Schmidt image with the same field of view
of the {\it GALEX} NUV image.  ({\it bottom panel}) UV F336W
(10\arcsec $\times$ 10\arcsec) image from HST-ACS which reveals the
complex system of dust within this galaxy.  The box in the NUV
and UKSTU images marks the HST-ACS FOV.
\label{fig3}}
\end{figure}
%--------------------------end Figure 3-------------------------------------
%--------------------------Figure 4----------------------------------
\begin{figure}
\includegraphics[width=6.7cm]{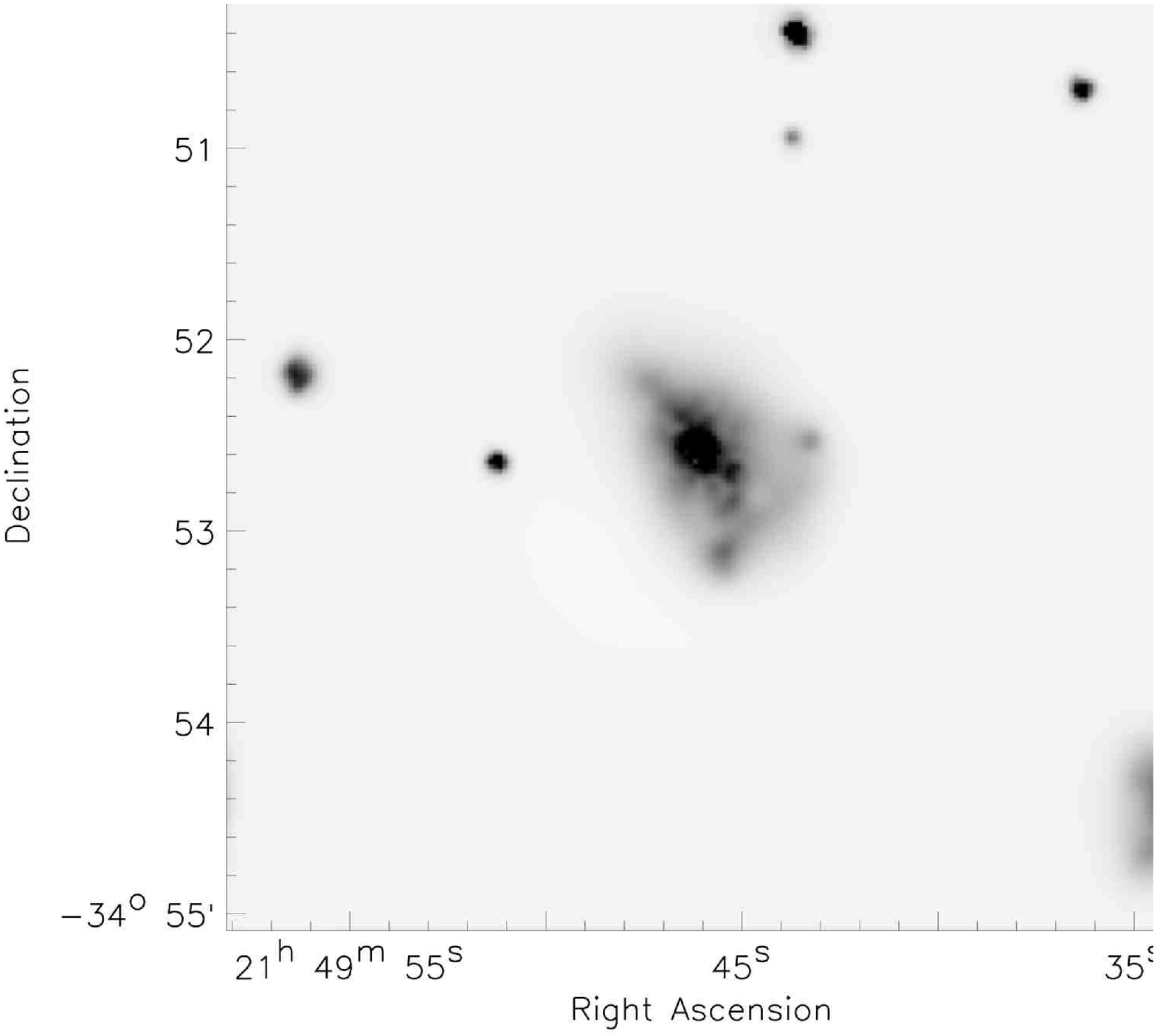}
\includegraphics[width=6.7cm]{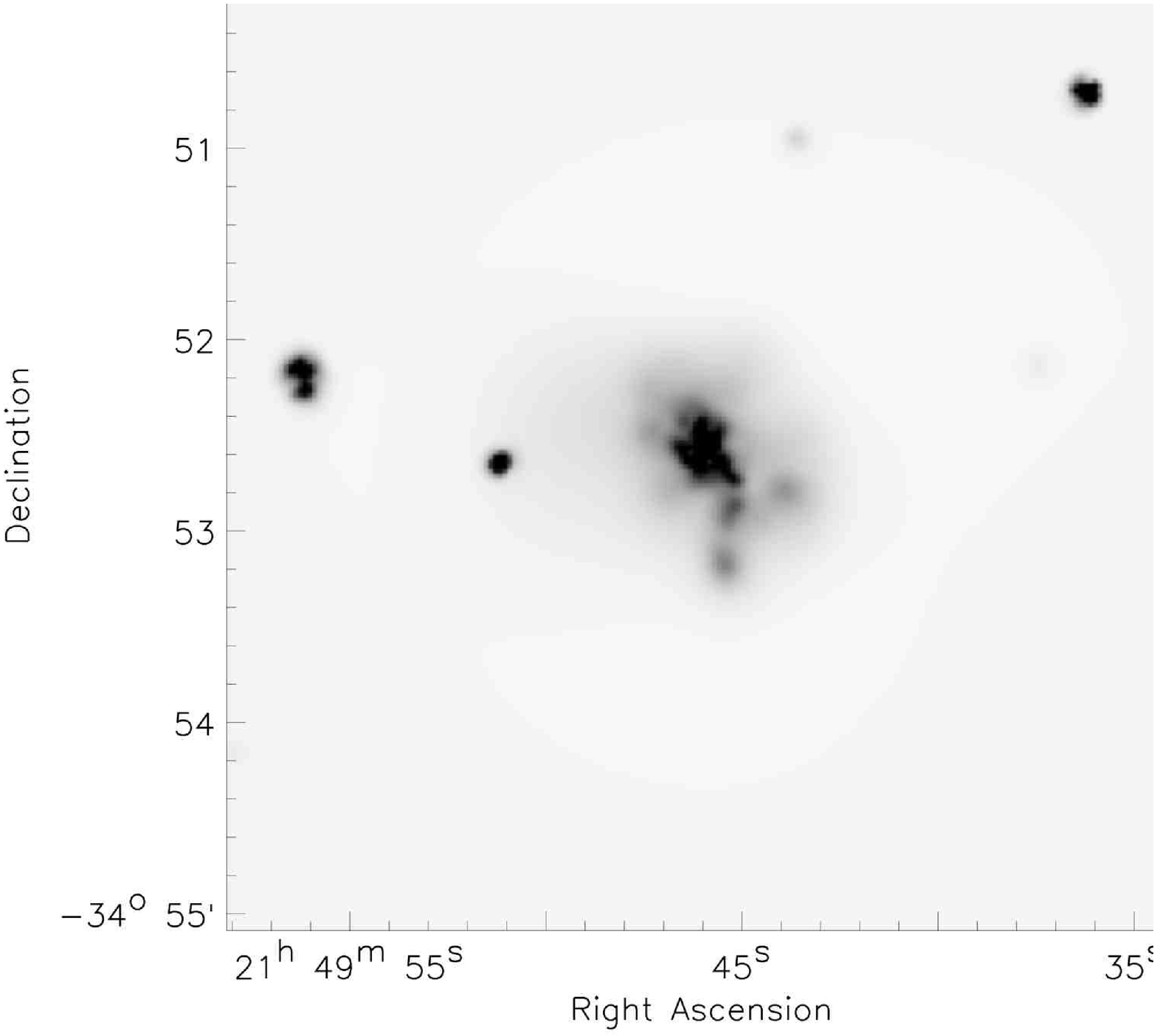}
\includegraphics[width=6.7cm]{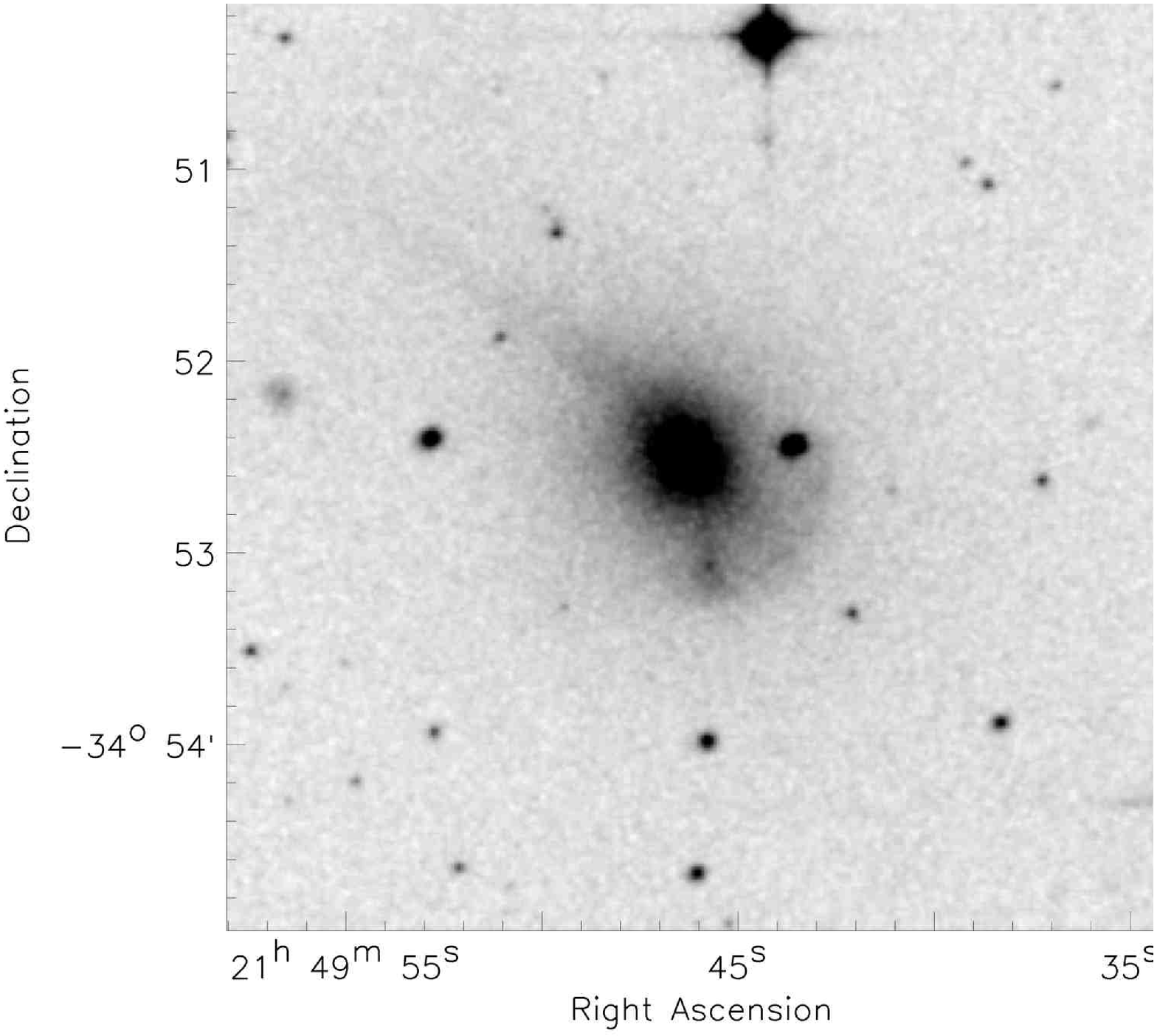}
\caption{{\it GALEX} UV and optical images  in a 5$\times$5 arcmin$^2$ field around
NGC 7135 galaxy.  {\it Top and mid panels}: NUV and FUV images
obtained with {\tt ASMOOTH} for $\tau_{min} =$8 and $\tau_{min} =$4
respectively.  The contrast of the FUV image has been enhanced in
order to show the extension of the emission. {\it Bottom panel}: UKSTU
Schmidt image.
\label{fig4}}
\end{figure}
%--------------------------end Figure 4-------------------------------------

 %--------------------------Figure 5----------------------------------
\begin{figure}
\includegraphics[width=7.4cm]{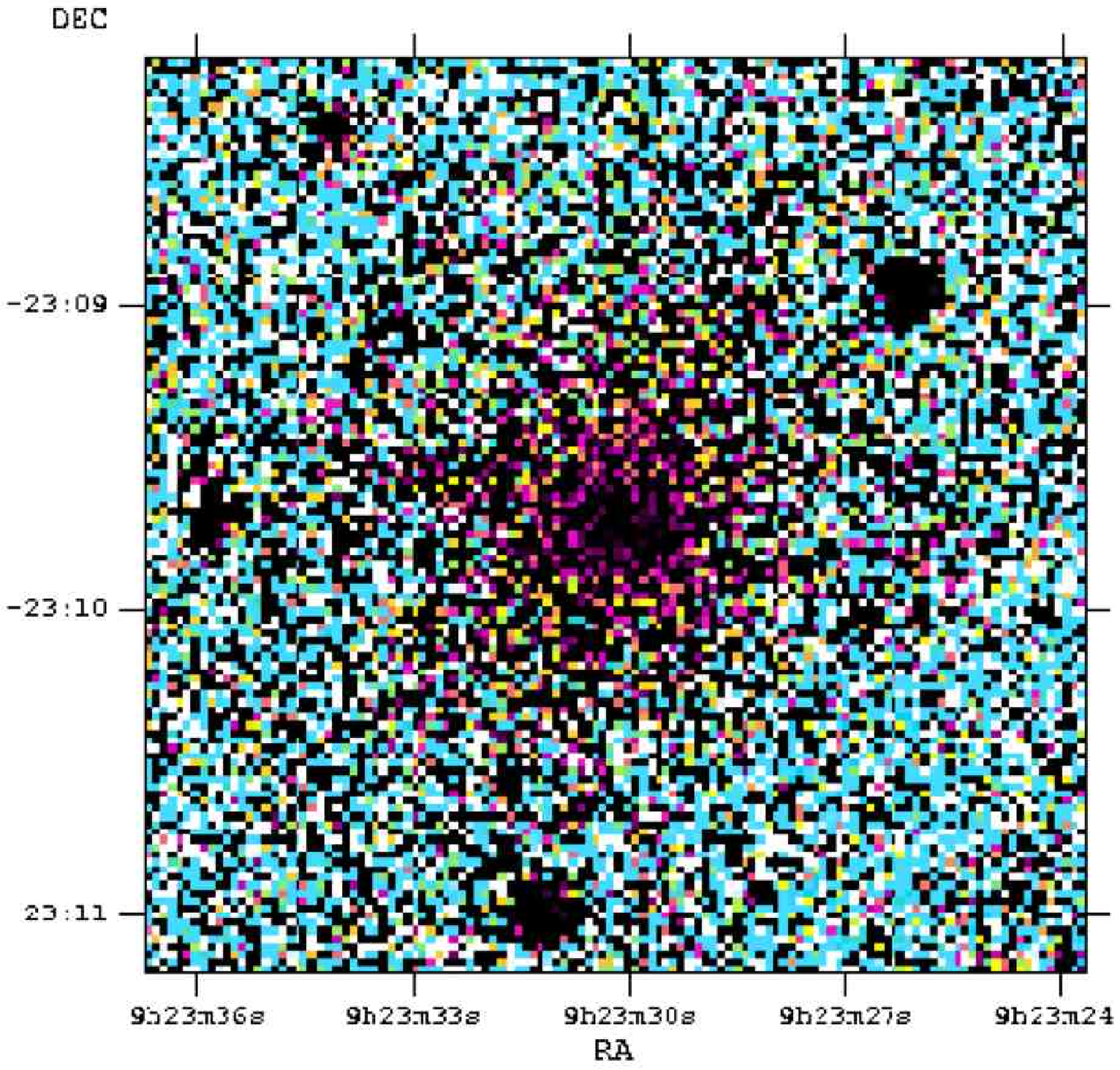}
\includegraphics[width=7.4cm]{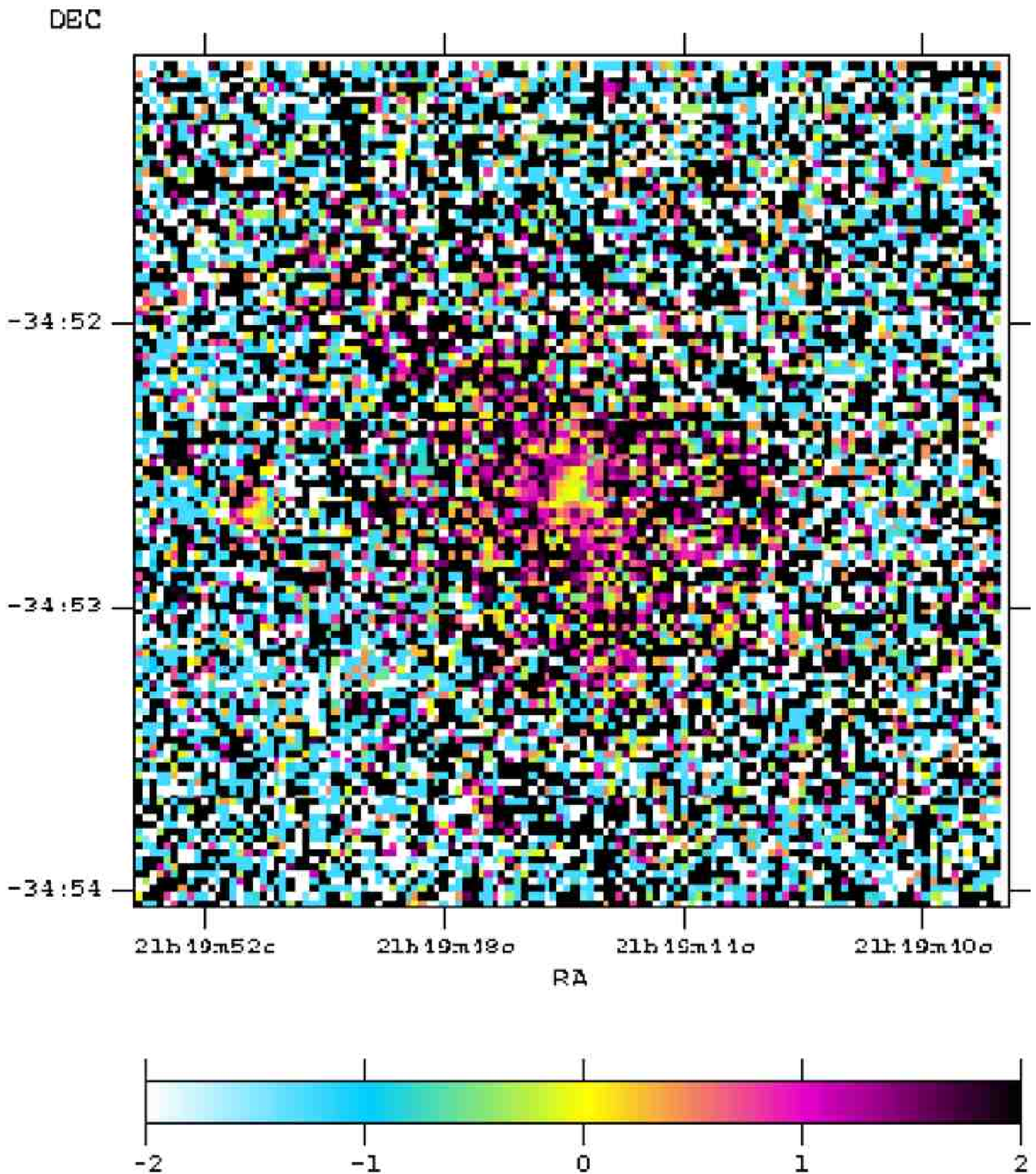}
\caption{ {\it GALEX} FUV-NUV 2D map of NGC~2865 {\it top panel})
and NGC 7135 ({\it bottom panel}). The colour-colour(FUV - NUV) maps
of NGC~2865 and NGC~7135 are derived using NUV and FUV images of
equivalent exposure time.
\label{fig5}}
\end{figure}
%--------------------------end Figure 5-------------------------------------

A chain of bright NUV sources in the East side of NGC~2865
(indicated with $3$ in the top panel of Figure~\ref{fig2}) has  a
counterpart in some optically faint galaxies, barely detectable in
the HST-ACS F606W image.  From West to East we identify the following
galaxies:
 (a) 9~23~33.4      -23~10~20.2 mag$_{NUV}$=20.86$\pm$0.07,
 (b) 9~23~34.2      -23~10~24.9, mag$_{NUV}$=23.01$\pm$0.18,
 (c) 9~23~35.1      -23~10~20.1, mag$_{NUV}$=22.88$\pm$0.16,
 (d) 9~23~35.7      -23~10~11.7, mag$_{NUV}$=22.21$\pm$0.13, and
 (e) 9~23~36.7      -23~10~11.3, mag$_{NUV}$=22.08$\pm$0.12. No
measurements of redshift are available for these galaxies.

\medskip
\noindent{\bf NGC~5018}

The galaxy is characterized by the presence of a complex dust--lane
system in the inner parts also visible in the HST-ACS F336W pass-band
which is shown in the bottom panel of Figure~\ref{fig3}.

In order to explore the galaxy fine structure we applied the adaptive
smoothing technique to the image. However, the low signal does not
allow us to reveal any detail of the shell structure sketched in the
optical by \citet{Fort86} (see their Figure~5).

\medskip
\noindent{\bf NGC~7135}

Contrary to the previous cases, the fine structures visible in the
optical DSS image are also revealed both in the NUV and FUV images
(see Figure~\ref{fig4} top and mid panels).  Specifically, the SW
shell as well as the faint plume extending from the galaxy center
towards the South are clearly visible.  In the NUV image, the
filamentary tidal tail in the NW direction of the galaxy nucleus
visible in the optical image is also seen. This feature makes NGC~7135
a possible analog of the ``Medusa'' galaxy, NGC 4194,
\citep[see][]{Rampazzo03}.

Both {\it GALEX} images  suggest the presence of dust in the galaxy
nucleus.

\subsection{NUV and FUV photometry}

Figure~\ref{fig5} shows the colour (FUV - NUV) maps of NGC~2865 and
NGC~7135. The NUV and FUV luminosity profiles of NGC~2865 and
NGC~7135 and the NUV profile of NGC~5018 are shown in
Figure~\ref{fig6}. Due to the low expousure time, we do not show the
luminosity profile of NGC~5018 in FUV because the S/N is very low.
We present also the F606W and F336W luminosity profiles of NGC~2865
and NGC~5018 respectively. The related colour profiles are plotted
in the bottom panels of Figure~\ref{fig6}.

NGC~2865 has a nearly flat and red colour profile with
(FUV-NUV)$\approx$2 throughout the galaxy. NGC~7135 is quite blue in
the center (FUV-NUV)$\approx$0 and becomes as red as
(FUV-NUV)$\approx$1.5 in the outskirts including the region where the
faint plume and the shell-like feature are present.

In Table~\ref{table3} we present the salient photometric data: the
effective radius r$_{e}$, i.e. the radius containing half of the total
galaxy light, the Sersic index, $n$, the total magnitude computed from
the surface photometry in the FUV and NUV bands and the global
(FUV-NUV) colour.

The r$_{e}$ values obtained in the NUV and the FUV bands for NGC~2865
are different. The two images have been obtained with different
exposure times (see Table~2). The better determined value of r$_{e}$
in the NUV band is consistent with that estimated from the F606W
HST-ACS optical image and in agreement with the the B-band values
reported in \citet{RC3}.

Both in NGC~2865 and NGC~7135, we measured different values of the
Sersic index $n$ in the NUV and FUV bands (see Table~3).  The fit is
very sensitive to the profile structure. The FUV and NUV luminosity
profiles of NGC~7135 appear to be really different throughout the
galaxy as measured by their very different Sersic indices $n$.  In the
F606W image the Sersic index we obtained for NGC~2865 is $n$=1.8
consistent both with NUV and FUV values although in the optical image
the presence of fine structures influences more strongly the
determination of the luminosity profile than in the NUV and FUV
images.

NGC~2865 and NGC~7135 show quite different (FUV-NUV)
gradients. Recent simulations of chemo-dynamical evolution of
elliptical galaxies (including radiative cooling, star formation,
feedback from Type II and Ia SNae, stellar winds and chemical
enrichment) performed by \citet{ChiosiCarraro2002},
 \citet{Kobayashi04}  and \citet{MerlinChiosi06} show strong
radial metallicity gradients in dissipation-less and monolithic
collapse of galaxy formation, respectively. Much shallower gradients
are foreseen if galaxies formed through major mergers. From an
observational point of view, both NGC 2865 and NGC 7135 colour
profiles are to some extent peculiar since ellipticals show, on the
average, a reddening of the colours towards the center
\citep{Peletier90} as a consequence of a metallicity gradient. On
the ground of the lack of correlation between both the
metallicity--$\sigma_c$ and [$\alpha$/Fe]--$\sigma_c$ relations
($\sigma_c$ and [$\alpha$/Fe] are central velocity dispersion and
the $\alpha$-- enhancement) \citet{Annibali07} suggest that the
young ages measured for some of their early-type galaxies are not
the result of a more prolonged star formation.  In particular, for
NGC 7135, they support the idea that the young age measured is due
to a recent rejuvenation episode of which the presence of the shells
morphology is a further indication. In the following sections we
will investigate if the above hypothesis is also supported by our
{\it GALEX} data.

 %----------------------------------------------Table 3 ------------------------
\begin{table}
\caption{{Summary of the \it GALEX} photometric data }
\begin{tabular}{lccc}
\hline \multicolumn{1}{l}{}& \multicolumn{1}{c}{NGC~2865} &
\multicolumn{1}{l}{NGC~5018}  &
\multicolumn{1}{l}{NGC~7135} \\
\hline
r$_{e}^{NUV}$ [arcsec]       &    27      &   34       &    52      \\
r$_{e}^{FUV}$ [arcsec]       &    37      &            &    52      \\
$n_{Sersic}$  [NUV]          &    1.5     &   3.3      &     2      \\
$n_{Sersic}$ [FUV]           &    2.2     &            &     4      \\
m$^{tot}_{FUV}$(1530\AA) & 18.59$\pm$0.14 &            & 17.85$\pm$0.25\\
m$^{tot}_{NUV}$(2310\AA) & 16.84$\pm$0.10 & 16.70$\pm$0.14 & 16.87$\pm$0.13  \\
(FUV-NUV)$^{tot}$            & 1.76$\pm$0.23  &            & 0.99$\pm$0.38  \\
\hline
\end{tabular}
\label{table3}
\end{table}
%--------------------------end Table 3--------------------------------

\section{Discussion}

\subsection{Shell galaxies and their environment}

The {\it GALEX} wide field of view allows us to study the environment
of shell galaxies. Table~\ref{table1} reports the low galaxy density
estimated by \citet{Tully88} for our objects. Given that shell
galaxies are typically found in low density environments, both their
high frequency (16.5\% of field Es show shells) and their most likely
interpretation (the result of a recent accretion/merging event)
suggest that these systems could trace the typical secular evolution
of early-type galaxies within poor, loose groups.

In this context, the S0 galaxy NGC~474 with a large shell system and
member of Arp 227 together with the spiral NGC~470, could be a typical
example. \citet{Rampazzo06} argued that both the large shell system
(likely due to an accretion episode) and the gas rich, faint late-type
companions detected in H\,{\sevensize I} south of NGC~470 would
suggest that Arp~227 is a ``poor group in evolution".  Therefore, it
could show us how the evolution proceeds in low density
environments. Indeed both the ($M_{H_2}/M_{H\,{\sevensize I} +H_2}$)
and L$\rm_X$/L$\rm_B$ ratios for NGC~470 are indistinguishable from
those of isolated galaxies while the X-ray emission of NGC~474 is
barely detected by XMM-Newton observations
\citet{Rampazzo06} at odds with evolved groups and/or so called
``fossil groups" \citep{Sansom00}. Arp~227 could be a snapshot of a
group in early evolutionary phases, the drivers of which are the
accretion of faint companions and the ongoing large scale
interaction between the dominant members NGC~470 and NGC~474.

%--------------------------Figure 6 ----------------------------------
\begin{figure*}
\includegraphics[width=10.0cm,angle=-90]{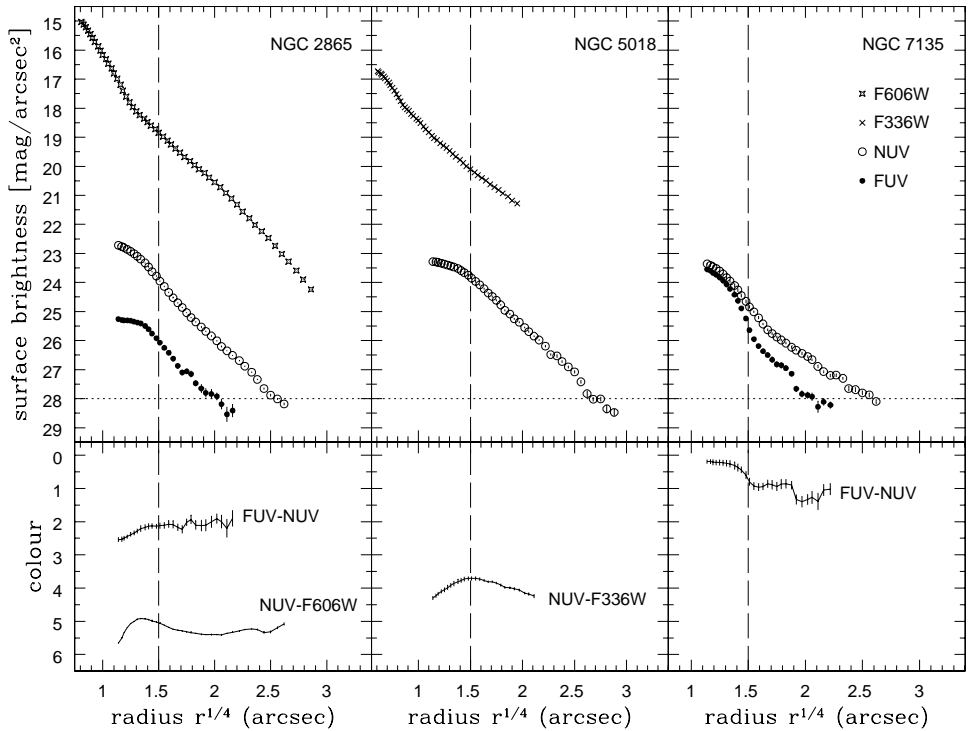}
\caption{({\it top left panel}) {\it GALEX} UV and optical surface
brightness profiles of NGC 2865.  ({\it top mid panel}) Near UV and
{\it GALEX} NUV luminosity profiles of NGC 5018 . ({\it top right
panel}) {\it GALEX} UV  surface brightness profile of NGC 7135.
({\it Bottom panels}): colours profiles from left to right of NGC
2865, NGC 5018 and NGC 7135. With the vertical dashed line at
5\arcsec\  the approximate FWHM of the GALEX PSF is indicated.
\label{fig6}}
\end{figure*}
%--------------------------end Figure 6----------------------------------

Following this  hypothesis we looked for  the large scale
distributions of H\,{\sevensize I}  available in the literature that
could be associated to our shell galaxies and to possible
companions. {\it GALEX} FUV and NUV data will enrich the picture
providing us with the star formation status of the member galaxies.

\begin{figure*}
{\includegraphics[width=15.0cm]{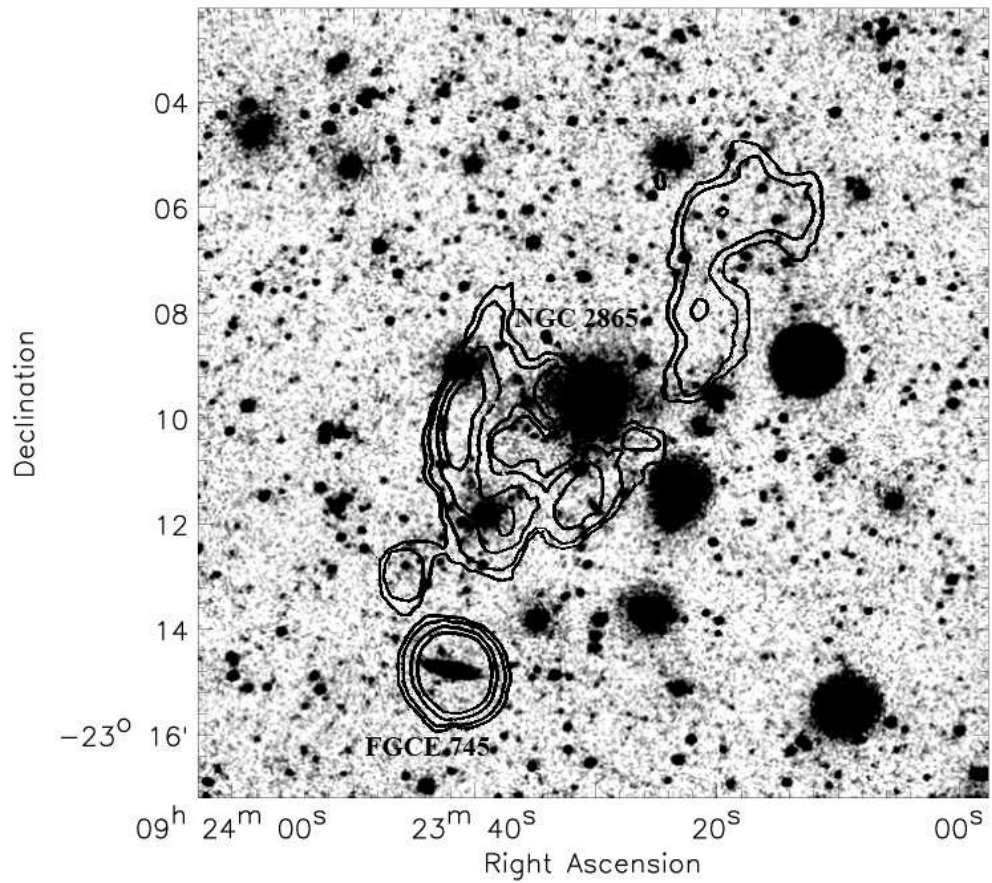}} \caption{NGC~2865.
VLA B+C -array, 20\arcsec\ resolution H\,{\sevensize I}
observations performed by \citet{Schiminovich95} overlaid  on the
{\it GALEX} NUV image. The FOV is
13\arcmin $\times$13\arcmin. The galaxy SE of NGC~2865 is FGCE~745.
\label{fig7}}
\end{figure*}

Figure~\ref{fig7} shows the overlay of the NGC~2865 H\,{\sevensize
I} observations  by \citet{Schiminovich95} with our {\it GALEX} NUV
image. The NUV image offers the same indications about the
association of fine structure and H\,{\sevensize I} emission as the
optical image. The atomic gas  seems to be associated with the
outermost east shell of NGC~2865, while a break is present in the
centre of the galaxy and in the position of the West protrusion.

Contrary to the case of the DSS image \citep[][see their Figure
1b]{Schiminovich95}, in our deep NUV image (Figure~\ref{fig7}) it is
only possible to guess both the large North West stellar loop
present in the optical image and the faint ribbon that projects
outwards to the South East. Along this latter we notice however an
arc-like increase of the NUV luminosity without any obvious optical
counterparts in the DSS image.

The H\,{\sevensize I} map shows that there is at least a companion
5.7\arcmin\ South-East of NGC~2865, namely FGCE~745 whose {\it
GALEX} magnitudes are mag$_{NUV}$=18.77$\pm$0.01 and
mag$_{FUV}$=19.34$\pm$0.04. The (FUV-NUV)=0.57$\pm$0.04 colour is
consistent with a late-type classification \citep{GildePaz06}. The
optical systemic velocity of NGC~2865 is 2627$\pm$3 km~s$^{-1}$. The
HI detection by \citet{Schiminovich95} indicates that the systemic
velocity of FGCE~745 is $\approx$2725 km~s$^{-1}$. More recently,
\citep{Mattews00} measured a systemic velocity of
2480$\pm$14  km~s$^{-1}$ for FGCE~745 suggesting that  the galaxy
is connected to the NGC~2865 South-Est H\,{\sevensize I} tail.

%\citet{Schiminovich95} measured the H\,{\sevensize I} emission coming
%also from an un-catalogued dwarf companion $\approx$9.3\arcmin\ west of NGC~2865.

%--------------------------Figure 7----------------------------------

%--------------------------end Figure 7-------------------------------------

%--------------------------Figure 8----------------------------------
\begin{figure*}
{\includegraphics[width=15.0cm]{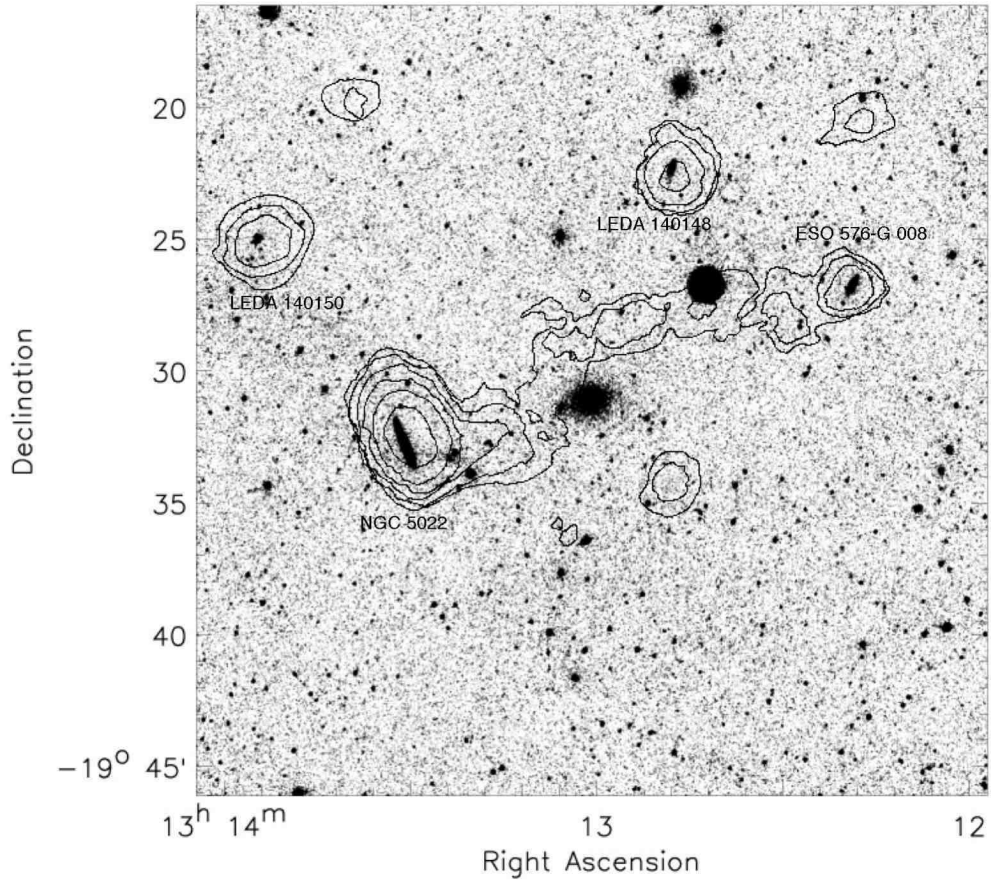}} \caption{NGC~5018. HI
observations overlaid  on the {\it GALEX} NUV original adaptively
smoothed  image. The FOV is 30\arcmin $\times$30\arcmin. The HI
image obtained with  VLA C+D-array, has a resolution of 60\arcsec\
\citep{Kim88}. We label in the figure the galaxies that HI reveals to
belong to the groups around NGC~5018. The radial velocity interval
of galaxies labeled in the figure, including NGC~5018 at the center
of the field, is $\Delta V \approx$ 420 km~s$^{-1}$.
\label{fig8}}
\end{figure*}
%--------------------------end Figure 8-------------------------------------

%--------------------------Figure 9----------------------------------
\begin{figure*}
{\includegraphics[width=15.0cm]{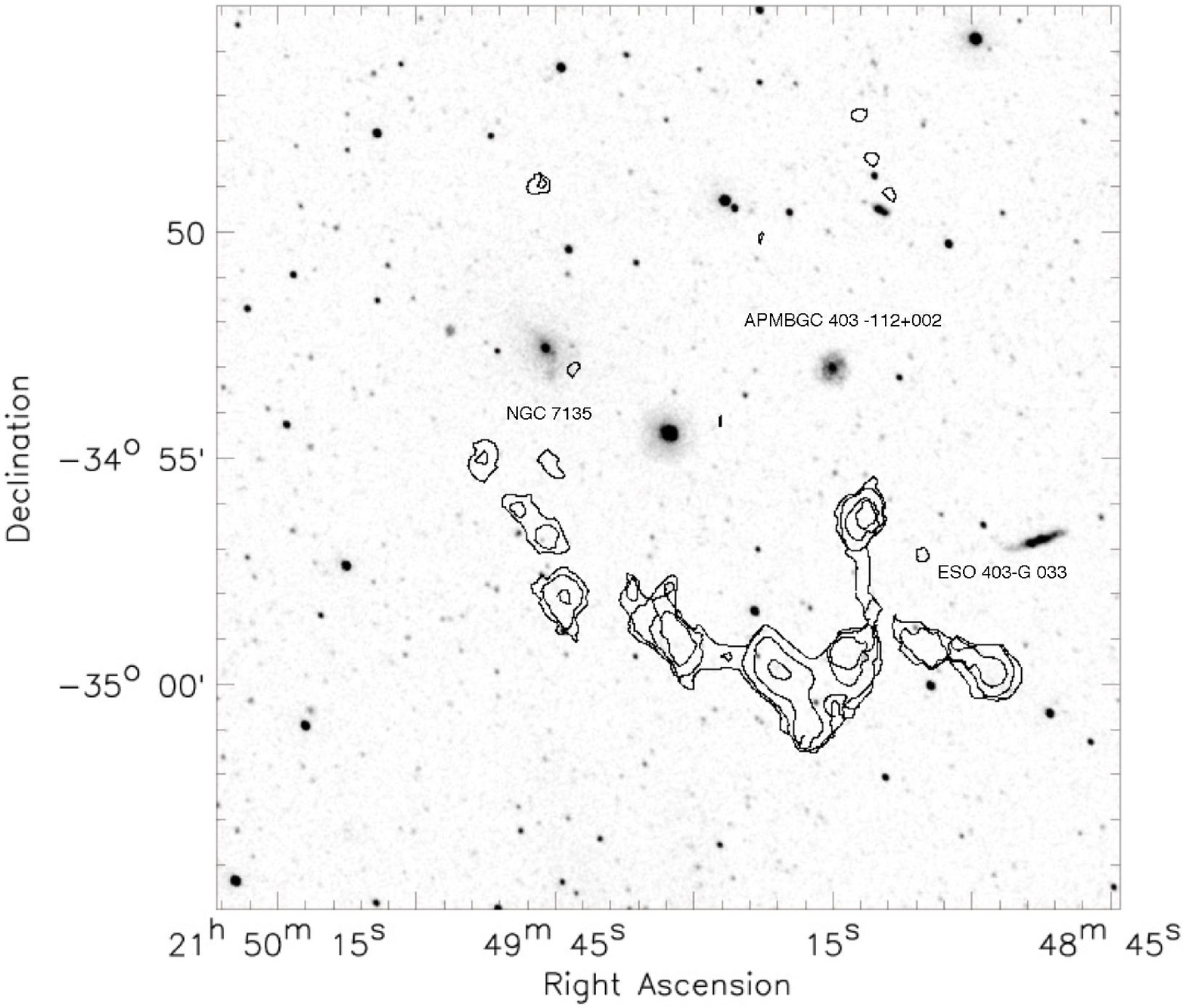}} \caption{NGC~7135. HI
observations  overlaid  on the {\it GALEX} NUV original adaptively
smoothed  image. The FOV is 20\arcmin $\times$20\arcmin. The HI
image obtained with  VLA C+D-array, has a resolution of
60.4\arcsec$\times$43.2\arcsec\ \citet{Schiminovich00}. Galaxies
West and South West of NGC~7135 are  APMBGC 403 -112+002 (4831
km~s$^{-1}$ from {\tt NED}, i.e. not bound to NGC~7135) and ESO
403-G 033 (redshift unknown) respectively.
\label{fig9}}
\end{figure*}
%--------------------------end Figure 9-------------------------------------

Also the environments  of NGC~5018 and NGC~7135 are quite rich in
cold gas. Figure~\ref{fig8} shows the H\,{\sevensize I} image of the
field containing NGC~5018 obtained with  VLA C+D-array by
\citet{Kim88}. There are four H\,{\sevensize I} rich galaxies
physically associated to this {\it bona fide} elliptical, namely
NGC~5022 (V$_{hel}$= 3001$\pm$8 km~s$^{-1}$;
mag$_{NUV}$=17.12$\pm$0.01), ESO~576-G 008 (V$_{hel}$=2691$\pm$39
km~s$^{-1}$; mag$_{NUV}$=18.10$\pm$0.02), LEDA~140148
(V$_{hel}$=3110 km~s$^{-1}$; mag$_{NUV}$=17.82$\pm$0.03), and
LEDA~140150 (V$_{hel}$=2732 km~s$^{-1}$;
mag$_{NUV}$=19.61$\pm$0.06). According to \citet{Kim88}, there is a
clear indication of interaction between NGC~5018 and the spiral
companion NGC~5022 from which the gas appears to be flowing towards
NGC~5018. The above authors report also about the presence of
several blobs of  H\,{\sevensize I}, which are marginally detected
within a few arcminutes from the optical centre of NGC~5018, and of
other three H\,{\sevensize I} features in the group which are not
associated with any visible galaxy.

\citet{Schiminovich00}  report the presence of a  puzzling,
large H\,{\sevensize I} filamentary structure nearby NGC~7135. In
the field of NGC~7135 shown by Figure~\ref{fig9} are present two
galaxies of comparable size:  APMUKS(BJ) B214615.65-350701.6 and
ESO~403-G 033 West and South West of the galaxy, respectively. The
first is not associated to NGC~7135 since it has V$_{hel}$=4831
km~s$^{-1}$. No redshift measures are available for the second
object. Neither the FUV nor the NUV images show objects possibly
associated to the H\,{\sevensize I} filamentary structure.

\subsection{Do shell galaxies host a recent starburst?}

\citet{Longhetti00} studied the H$\beta$ vs. [MgFe] plane trying to
infer the age of the stellar population in shell galaxies comparing
their line--strength indices with those of a sample of interacting and
{\it normal} early-type galaxies. In the above plane, shell galaxies
show the same distribution of {\it normal} early-type, although a
number of shell galaxies and pairs have a much stronger
H$\beta$. \citet{Longhetti00} suggest that the scatter in the H$\beta$
vs. [MgFe] plane could be due to a common origin, perhaps a secondary
episode of star formation after their formation. The explanation comes
naturally for shell (and pair galaxies) where the signatures of
interaction are evident. The shape of the distribution of the
line-strength indices in the H$\beta$ vs. [MgFe] plane suggests the
presence of an effect of a metal enrichment that always accompany star
formation. \citet{Longhetti99} further suggest that, if the last star
forming event is connected to the formation of the shell system as
expected from simulations, the shells ought to be a long lasting
phenomenon since star forming events that occurred in the nuclear
region of shell galaxies are statistically old (from 0.1 up to several
Gyr). It is worth mentioning here the study by \citet{Tantalo04b} who
added another dimension to the problem, i.e.  the effect of the degree
of enhancement in $\alpha$ elements in the abundance patterns. They
argued that part of the scatter in H$\beta$ could reflect the
intrinsic variation of $\alpha$ enhancement from galaxy to galaxy
existing in old populations of stars due to different star formation
histories rather than a dispersion in age caused by more recent star
forming episodes.

Based on these premises, we investigate whether the {\it GALEX} data
may confirm or disprove the case that shell galaxies hosted a recent
star burst event. Given that UV colours are considered to be
particularly suited to identify very young stellar populations and
their distribution inside interacting galaxies \citep[see
e.g.][]{Hibbard05}, to (FUV-NUV) we also add the UV-optical
colours. There is, however, an important point to keep in mind dealing
with UV colours, i.e. the possibility that high metallicity, old stars
(age $\geq 10^{10}$ yr and metallicity $Z\simeq 3Z_\odot$) likely
present even in small percentages in early-type galaxies could affect
the (FUV-NUV) colours. The evolved stars in metal rich populations
(Extremely HB and AGB-manqu\'e) are the proposed source of the UV
excess in quiescent early-type galaxies \citep{Bressan94} and any
modeling of star formation must take this population into
consideration.  See \citet{Bressan94} for a detailed discussion of the
subject. Fortunately, we can consider additional clues (presence of
dust and gas, dynamics, overall spectral energy distribution) in
interpreting the results.

Our shell galaxies are {\it dynamically young} objects and belong to
environments containing fresh gas in the ionized, neutral and
molecular forms. However, the spatial distribution of the gas changes
from galaxy to galaxy. Only for NGC~7135 is there some indication in
the literature for ongoing gas re-fuelling. In the centre of this
galaxy \citet{Rampazzo03} detected in the H$\alpha$ emission line the
presence of an ionized gaseous component for a total extension of
$\approx$20\arcsec, elongated in the southwest direction in
correspondence to the FUV emission (Figure~\ref{fig4} mid panel). The
ionized gas and the stars differ significantly in systemic velocity
(see \S~2) suggesting that the gas likely originates from an
interaction/accretion episode suffered by the galaxy.

For our three galaxies estimates of {\bf {\it the mean luminosity
weighted ages}} have been derived from optical line-strength indices
by \citet{Longhetti99}, \citet{Leonardi00}, \citet{Longhetti00},
\citet{Rampazzo05}, and \citet{Annibali07}. For comparison we
consider  here the set of early-type galaxies analyzed by
\citet{Rampazzo05}, \citet{Annibali06}, and \citet{Annibali07}.
These galaxies are characterized  by (i)  the presence of ionized
gas, (ii) their location in low density environments, also typical
of shell galaxies, and (iii)  the presence of
dynamical/morphological peculiarities in a high percentage of them.
Finally, there are also a few additional shell systems.  In addition
to the three galaxies of this study, out of the total  sample of 65
objects studied by \citet{Rampazzo05}, \citet{Annibali06}, and
\citet{Annibali07},  we have taken 15 early-type galaxies for which
the {\it GALEX} measurements are already to disposal. In total we
consider a sample of 18 galaxies. All relevant data are summarized
in  Tables~\ref{table4} and \ref{table5}.

%%%%%%%%%%%%%%%Table 4
\begin{table*}
%\begin{center}
\caption{Data for the sample of 18 galaxies examined in this study.
Columns (1) through (8) list the galaxy identification, the total
FUV, NUV and V magnitudes, the total central colour (B-V)$_{tc}$,
the central velocity dispersion $\sigma_c$ in km/s, the effective
radius $r_e$ in arcsec, and the colour (FUV-NUV) within $r_e/8$.}
\begin{tabular}{lllrllll}
\hline \multicolumn{1}{l}{Galaxy}& \multicolumn{1}{l}{NUV}&
\multicolumn{1}{l}{FUV} & \multicolumn{1}{l}{V} &
\multicolumn{1}{l}{(B-V)$_{tc}$} & \multicolumn{1}{l}{log
$\sigma_c$} & \multicolumn{1}{l}{$r_{e}$} &
\multicolumn{1}{l}{(FUV-NUV)} \\
\multicolumn{1}{l}{ident.}& \multicolumn{1}{l}{total}&
\multicolumn{1}{l}{total} & \multicolumn{1}{l}{total} &
\multicolumn{1}{l}{} & \multicolumn{1}{l}{[km~s$^{-1}$]} &
\multicolumn{1}{l}{[arcsec]} &
\multicolumn{1}{l}{[$_{r_{e}/8}$]} \\
\hline
NGC ~777 &17.19$\pm$0.01 & 17.70$\pm$0.02 &11.26$\pm$0.14 & 0.87$\pm$0.02 & 2.51$\pm$0.01 & 34.4 & 0.20$\pm$0.23 \\
NGC 1052 &15.83$\pm$0.01 & 16.85$\pm$0.01 &10.34$\pm$0.13 & 0.79$\pm$0.01 & 2.32$\pm$0.01 & 33.7 & 0.41$\pm$0.22\\
NGC 1380 &15.29$\pm$0.01 & 16.78$\pm$0.01 & 9.83$\pm$0.10 & 0.80$\pm$0.01 & 2.35$\pm$0.01 & 20.3 & 0.83$\pm$0.29 \\
NGC 1389 &16.78$\pm$0.01 & 18.63$\pm$0.02 &11.42$\pm$0.13 & 0.79$\pm$0.01 & 2.14$\pm$0.03 & 15.0 & 1.37$\pm$0.22\\
NGC 1407 &14.93$\pm$0.01 & 15.64$\pm$0.01 & 9.40$\pm$0.20 & 0.84$\pm$0.01 & 2.44$\pm$0.01 & 70.3 & 0.19$\pm$0.22 \\
NGC 1426 &16.84$\pm$0.03 & 18.30$\pm$0.08 &11.29$\pm$0.05 & 0.76$\pm$0.01 & 2.19$\pm$0.01 & 25.0 & 1.02$\pm$0.59\\
NGC 1453 &17.03$\pm$0.06 & 18.16$\pm$0.12 &11.13$\pm$0.13 & 0.83$\pm$0.01 & 2.48$\pm$0.03 & 25.0 & 0.80$\pm$1.20\\
NGC 1521 &17.15$\pm$0.33 & 18.66$\pm$0.11 &11.24$\pm$0.06 & 0.81$\pm$0.01 & 2.38$\pm$0.02 & 25.5 & 0.15$\pm$0.49\\
NGC 1553 &14.53$\pm$0.01 & 16.18$\pm$0.01 & 9.31$\pm$0.08 & 0.75$\pm$0.01 & 2.25$\pm$0.01 & 65.6 & 1.24$\pm$0.22\\
NGC 2865 &16.93$\pm$0.01 & 19.15$\pm$0.04 &11.34$\pm$0.14 & 0.71$\pm$0.01 & 2.26$\pm$0.02 & 12.5 & 2.52$\pm$0.28\\
NGC 4374 &14.35$\pm$0.01 & 15.60$\pm$0.01 & 8.93$\pm$0.05 & 0.82$\pm$0.01 & 2.45$\pm$0.01 & 50.9 & 0.59$\pm$0.22\\
NGC 4552 &14.83$\pm$0.01 & 15.73$\pm$0.01 & 9.57$\pm$0.05 & 0.82$\pm$0.01 & 2.42$\pm$0.01 & 29.3 & -0.05$\pm$0.22\\
NGC 5018 &16.45$\pm$0.08 &                &10.40$\pm$0.13 & 0.71$\pm$0.01 & 2.38$\pm$0.02 & 22.8 & 1.05$\pm$0.75 \\
NGC 5638 &16.27$\pm$0.01 & 17.92$\pm$0.02 &11.04$\pm$0.14 & 0.79$\pm$0.01 & 2.22$\pm$0.01 & 28.0 & 0.86$\pm$0.29 \\
NGC 5813 &15.99$\pm$0.01 & 17.10$\pm$0.02 &10.23$\pm$0.13 & 0.81$\pm$0.01 & 2.38$\pm$0.01 & 57.2 & 0.77$\pm$0.25 \\
NGC 6958 &16.42$\pm$0.01 & 18.40$\pm$0.04 &11.22$\pm$0.13 & 0.75$\pm$0.01 & 2.28$\pm$0.02 & 19.8 & 1.51$\pm$0.27 \\
NGC 7135 &17.02$\pm$0.01 & 18.44$\pm$0.02 &11.58$\pm$0.10 & 0.84$\pm$0.01 & 2.36$\pm$0.02 & 31.4 & -0.03$\pm$0.23\\
IC ~1459 &15.58$\pm$0.02 & 16.49$\pm$0.04 & 9.89$\pm$0.15 & 0.85$\pm$0.01 & 2.49$\pm$0.01 & 34.4 & 0.43$\pm$0.22\\
\hline
\end{tabular}
%\end{center}
\label{table4}
\begin{minipage}{0.7 \textwidth}
\footnotesize{The values of (FUV-NUV)  have been corrected for
foreground extinction \citep{Schlegel98}.}
\end{minipage}
\end{table*}

%%%%%%%%%%%%%%%%%%%Table 5
\begin{table*}
%\begin{center}
\caption{Line-strength indices within $r_e/8$ and age and
metallicity assignments for the sample of 18 galaxies listed in
Table~\ref{table4}. }
\begin{tabular}{lllllrl}
\hline
\multicolumn{1}{l}{Galaxy}& \multicolumn{1}{l}{H$\beta$} &
\multicolumn{1}{l}{H$\gamma$A} & \multicolumn{1}{l}{H$\delta$A} &
\multicolumn{1}{l}{Mg2} & \multicolumn{1}{l}{Age} &
\multicolumn{1}{l}{Z} \\
\multicolumn{1}{l}{ident.}& \multicolumn{1}{l}{[$_{r_{e}/8}$]} &
\multicolumn{1}{l}{[$_{r_{e}/8}$]} &
\multicolumn{1}{l}{[$_{r_{e}/8}$]} &
\multicolumn{1}{l}{[$_{r_{e}/8}$]} & \multicolumn{1}{l}{[Gyr]} &
\multicolumn{1}{l}{} \\
\hline
NGC ~777 & 1.48$\pm$0.80 & -5.340$\pm$0.803 & -3.080$\pm$0.585 & 0.336$\pm$0.013 & 5.4$\pm$2.1 & 0.024$\pm$0.004 \\
NGC 1052 & 3.21$\pm$0.09 & -7.960$\pm$0.143 & -3.250$\pm$0.109 & 0.335$\pm$0.003 &14.5$\pm$4.2 & 0.032$\pm$0.007 \\
NGC 1380 & 1.75$\pm$0.09 & -5.570$\pm$0.111 & -2.740$\pm$0.102 & 0.306$\pm$0.002 & 4.4$\pm$0.7 & 0.038$\pm$0.006 \\
NGC 1389 & 1.80$\pm$0.10 & -5.490$\pm$0.127 & -2.230$\pm$0.113 & 0.286$\pm$0.003 & 4.5$\pm$0.6 & 0.032$\pm$0.005 \\
NGC 1407 & 1.62$\pm$0.11 & -6.290$\pm$0.139 & -3.330$\pm$0.132 & 0.345$\pm$0.003 & 8.8$\pm$1.5 & 0.033$\pm$0.005 \\
NGC 1426 & 1.64$\pm$0.09 & -5.810$\pm$0.129 & -3.010$\pm$0.116 & 0.285$\pm$0.003 & 9.0$\pm$2.5 & 0.024$\pm$0.005 \\
NGC 1453 & 1.59$\pm$0.10 & -5.840$\pm$0.136 & -3.510$\pm$0.122 & 0.326$\pm$0.003 & 9.4$\pm$2.1 & 0.033$\pm$0.007 \\
NGC 1521 & 1.67$\pm$0.10 & -5.210$\pm$0.144 & -2.460$\pm$0.122 & 0.284$\pm$0.003 & 3.2$\pm$0.4 & 0.037$\pm$0.006 \\
NGC 1553 & 1.90$\pm$0.09 & -5.790$\pm$0.112 & -2.780$\pm$0.948 & 0.287$\pm$0.002 & 4.8$\pm$0.7 & 0.031$\pm$0.004 \\
NGC 2865 & 3.12$\pm$0.15 &                  &                  & 0.209$\pm$0.003 & 1.8$\pm$0.5 & \\
NGC 4374 & 1.75$\pm$0.16 & -6.170$\pm$0.226 & -2.670$\pm$0.129 & 0.314$\pm$0.005 & 9.8$\pm$3.4 & 0.025$\pm$0.010 \\
NGC 4552 & 1.29$\pm$0.13 & -6.550$\pm$0.152 & -3.180$\pm$0.155 & 0.341$\pm$0.004 & 6.0$\pm$1.4 & 0.043$\pm$0.012 \\
NGC 5018 & 2.68$\pm$0.15 &                  &                  & 0.209$\pm$0.003 & 2.8$\pm$0.0 &  \\
NGC 5638 & 1.62$\pm$0.14 & -6.190$\pm$0.203 &-3.010$\pm$0.170  & 0.308$\pm$0.005 & 9.1$\pm$2.3 & 0.024$\pm$0.008 \\
NGC 5813 & 1.78$\pm$0.09 & -6.280$\pm$0.119 &-3.070$\pm$0.110  & 0.301$\pm$0.002 &11.7$\pm$1.6 & 0.018$\pm$0.002 \\
NGC 6958 & 2.17$\pm$0.09 & -4.190$\pm$0.121 &-0.997$\pm$0.110  & 0.258$\pm$0.003 & 3.0$\pm$0.3 & 0.038$\pm$0.006 \\
NGC 7135 & 2.41$\pm$0.07 & -6.390$\pm$0.100 &-3.630$\pm$0.084  & 0.292$\pm$0.002 & 2.4$\pm$0.4 & 0.047$\pm$0.010 \\
IC ~1459 & 1.70$\pm$0.10 & -6.900$\pm$0.121 &-3.630$\pm$0.120  & 0.343$\pm$0.003 & 8.0$\pm$2.2 & 0.042$\pm$0.009 \\
\hline
\end{tabular}
%\end{center}
\label{table5}
\begin{minipage}{0.9 \textwidth}
\footnotesize{The age of NGC 2865 is derived from the analysis of
H$\beta$ and [CaII] indices by \citet{Longhetti99} and
\cite{Longhetti00}. For NGC 5018 we adopt the age estimated  by
\citet{Leonardi00} on the basis of [CaII] and
H$\delta$/$\lambda$4005 line-strength indices. For both galaxies the
H$\beta$ and Mg2 line-strength indices are taken from
\citet{Longhetti00}. }
\end{minipage}
\end{table*}

Table~\ref{table4} lists the total FUV, NUV and V magnitudes, the
colour (B-V), the central velocity dispersion $\sigma_c$ in km/s, the
effective radius $r_e$ in arcsec, and the colour (FUV-NUV)$_{r_e/8}$
(see below for the meaning of the suffix). Table~\ref{table5} lists
the values of line-strength indices that are customarily considered to
be more sensible either to the age (e.g.  H$\beta$, H$\gamma$,
H$\delta$) or the metallicity (e.g. Mg2). They are measured within an
aperture centered on the nucleus of the galaxy with radius r$_{e}$/8
(the effective radius, r$_{e}$, contains half of the total galaxy
light calculated in the B-band).  Finally we list the age and
metallicity assigned to each galaxy by
\citet{Rampazzo05}, \citet{Annibali06}, and \citet{Annibali07} as
the most likely solution in the (age, Z, and [$\alpha$/Fe]) space.

There is an important consideration to be made about the
``sensitivity" of the indices to the three most important parameters
(age, metallicity, and degree of $\alpha$-enhancement). The subject
has been recently addressed by \citet{Tantalo04b} from a quantitative
point of view. Limiting the discussion to H$\beta$ and Mg2, the
situation is as follows: H$\beta$ strongly depends on all the three
parameters with little resolving power; Mg2 primarily depends on
metallicity and age, little on the the degree of enhancement. For
details see Table 5 of \citet{Tantalo04b}. This may explain part of
the difficulties we have encountered with some galaxies in the
discussion below.

Using these data we look for significant correlations between the {\it
GALEX} (FUV-NUV)$_{r_e/8}$ colour measured within r$_{e}$/8 and the
line-strength indices and compare them with theoretical
predictions. To this aim we need to generate integrated magnitudes and
colours of Single Stellar Populations (SSPs) with different chemical
compositions in the GALEX pass-bands.

\subsection{Modelling GALEX magnitudes and colours}\label{model}

In the following we briefly summarize the key assumptions and features
of the stellar models and isochrones, library of stellar spectra and
the ABmag photometric system that are used to calculate theoretical
FUV and NUV magnitudes for SSPs.

\begin{description}
\item [{\bf Stellar models and isochrones.}]
We have adopted the Padova Library of stellar models and companion
isochrones according to the release by \citet{Girardi00} and Girardi
(2003, private communication). This particular set of stellar
models/isochrones differs from the classical one by
\citet{Bertelli94} for the efficiency of convective overshooting and
the prescription for the mass-loss rate along the Asymptotic
Giant Branch (AGB) phase. The stellar models extend from the ZAMS up
to either the start of the thermally pulsing AGB phase (TP-AGB) or
carbon ignition. No details on the stellar models are given here;
they can be found in \citet{Girardi00} and \citet{Girardi02}.
Suffice it to mention that: (i) in low mass stars passing from the
tip of red giant branch (T-RGB) to the HB or clump, mass-loss by
stellar winds is included according to the \citet{Reimers75} rate
with $\eta$=0.45; (ii) the whole TP-AGB phase is included in the
isochrones with ages older than 0.1\,Gyr according to the algorithm
of \citet{Girardi98} and the mass-loss rate of
\citet{Vassiliadis93}; (iii) four chemical compositions are
considered as listed in Table~\ref{enh-deg}.\\

\item[{\bf Library of stellar spectra.}]
The library of stellar spectra is taken from \citet{Girardi02}. It
covers a large range of the $\log{\Teff}$ -- $\log{\rm g}$ -- and
[M/H] space. No details are given here. Suffice to mention

\begin{itemize}
\item The basic spectra are from Kurucz ATLAS9 non-overshooting models
\citep{Castelli97,Bessell98} complemented with:

\item Black-body spectra for \Teff$>$50,000\,K;

\item \citet{Fluks94} empirical M-giant spectra, extended with synthetic
ones in the IR and UV, and modified short-ward of 4000\AA\ so as to
produce reasonable \Teff-$(U-B)$ and \Teff-$(B-V)$ relations for
cool giants;

\item \citet{Allard00} DUSTY99 synthetic spectra for M, L and T dwarfs.
\end{itemize}

Using the above library of stellar models and spectra, we have
calculated the SSPs integrated spectra following the method described
in \citet{Bressan94}.

It is worth recalling here that above a certain value of the
metallicity, say $Z=3\, Z_\odot$, these SSP exhibit the UV excess
caused by the so-called AGB manqu\'e stars of high metal content,
whose percentage in the stellar population mix of a galaxy is small
but high enough to give rise to the UV-upturn \citep[see][ for all
details]{Bressan94}.

The theoretical absorption-line indices in the Lick system have been
calculated as described in \citet[Sect.4]{Tantalo04a,Tantalo04b} with
the aid of the fitting functions ($\mathcal{FF}$s) given by
\citet{Worthey94} and \citet{Worthey97}.

The theoretical magnitudes used in this study have been calculated in
the ABmag system \citep[see][]{Oke74} for the Johnson-Cousins-Glass
UBVRIJHK system, using filter response curves from \citet{Bessell88}
and \citet{Bessell90} and for the two bands (FUV and NUV) of
GALEX\footnote{The whole library of SSP models can be found on the
website {\it http://dipastro.pd.astro.it/galadriel/}} as described
below.

\begin{table}
\small
\begin{center}
\caption[]{Chemical composition for the SSPs in use.}
\label{enh-deg}
\begin{tabular*}{34mm}{|c c c|}
\hline \multicolumn{1}{|c}{Z} & \multicolumn{1}{c}{Y} &
\multicolumn{1}{c|}{X} \\
\hline
 0.008& 0.248 & 0.7440 \\
 0.019& 0.273 & 0.7080 \\
 0.040& 0.320 & 0.6400 \\
 0.070& 0.338 & 0.5430 \\
\hline
\end{tabular*}
\end{center}
\end{table}

\item{\bf The ABmag system.} \label{absys}
If the spectral flux as it arrives at the Earth, $f_{\lambda}$, is
known, the apparent magnitude $m_{S_{\lambda}}$, in a given pass-band
with transmission curve $S_{\lambda}$ in the interval [$\lambda_{1}$,
$\lambda_{2}$], can be calculated by:

\begin{equation}
m_{S_{\lambda}} = -2.5 \log \left(
                   \frac{\int^{\lambda_{2}}_{\lambda_{2}} \lambda f_{\lambda}     S_{\lambda} d\lambda}
                        {\int^{\lambda_{2}}_{\lambda_{2}} \lambda f^{0}_{\lambda} S_{\lambda} d\lambda} \right) + m^0_{S_{\lambda}}~.
\label{magnitude}
\end{equation}

\noindent where $f^{0}_{\lambda}$ represents a reference spectrum
(not necessary a stellar one) that produces a known apparent
magnitude $m^{0}_{S_{\lambda}}$ Here, both $f^{0}_{\lambda}$ and
$m^{0}_{S_{\lambda}}$, completely define the ``zero-points'' of a
synthetic photometric system.

By photometric ``zero-points'', one usually means the constant
quantities that one should add to instrumental magnitude in order to
transform them to standard magnitudes, for each filter
$S_{\lambda}$. In the formalism adopted here, we do not make use of
the concept of instrumental magnitude, and hence such constants do
not need to be defined, and we consider as ``zero-points'' the
quantities in eqn.(\ref{magnitude}) that depend only on the choice
of $f^{0}_{\lambda}$ and $m^{0}_{S_{\lambda}}$, which are constant
for each filter.

In the original work by \citet{Oke64}, monochromatic AB magnitudes
are defined by

\begin{equation}
m_{AB,\nu} = -2.5 \log f_{\nu} - 48.60~.
\end{equation}

\noindent This means that a reference spectrum of constant flux
density per unit frequency

\begin{equation}
f_{AB,\nu}^{0} = 3.361\times10^{-20} [erg s^{-1} cm^{-2} Hz^{-1}]
\label{ref_spec}
\end{equation}

\noindent
will have AB magnitudes $m^{0}_{AB,\nu}$= 0 at all frequencies $\nu$.\\
This definition can be extended to any filter system, provided that
we replace the monochromatic flux $f_{\nu}$ with the photon counts
over each pass-band $S_{\lambda}$ obtained from the star, compared
to the photon counts that one would get by observing
$f^{0}_{AB,\nu}$:

\begin{equation}
m_{AB,S_{\lambda}} = -2.5 \log \left[
                     \frac{\int^{\lambda_{2}}_{\lambda_{2}} (\lambda / hc) f_{\lambda}        S_{\lambda} d\lambda}
                          {\int^{\lambda_{2}}_{\lambda_{2}} (\lambda / hc) f^{0}_{AB,\lambda} S_{\lambda} d\lambda} \right]~,
\label{mab}
\end{equation}

\noindent where $f^{0}_{AB,\lambda}$=$f^{0}_{AB,\nu} c/\lambda^{2}$.
It is easy to show that eqn.(\ref{mab}) is just a particular case of
the most general eqn.(\ref{magnitude}) for which $f^{0}_{AB,\nu}$ is
the reference spectrum given by eqn.(\ref{ref_spec}), and
$m^{0}_{AB,S_{\lambda}}$= 0 are the reference magnitudes.

\end{description}

%--------------------------Figure 10----------------------------------
\begin{figure*}
\includegraphics[width=16.8cm]{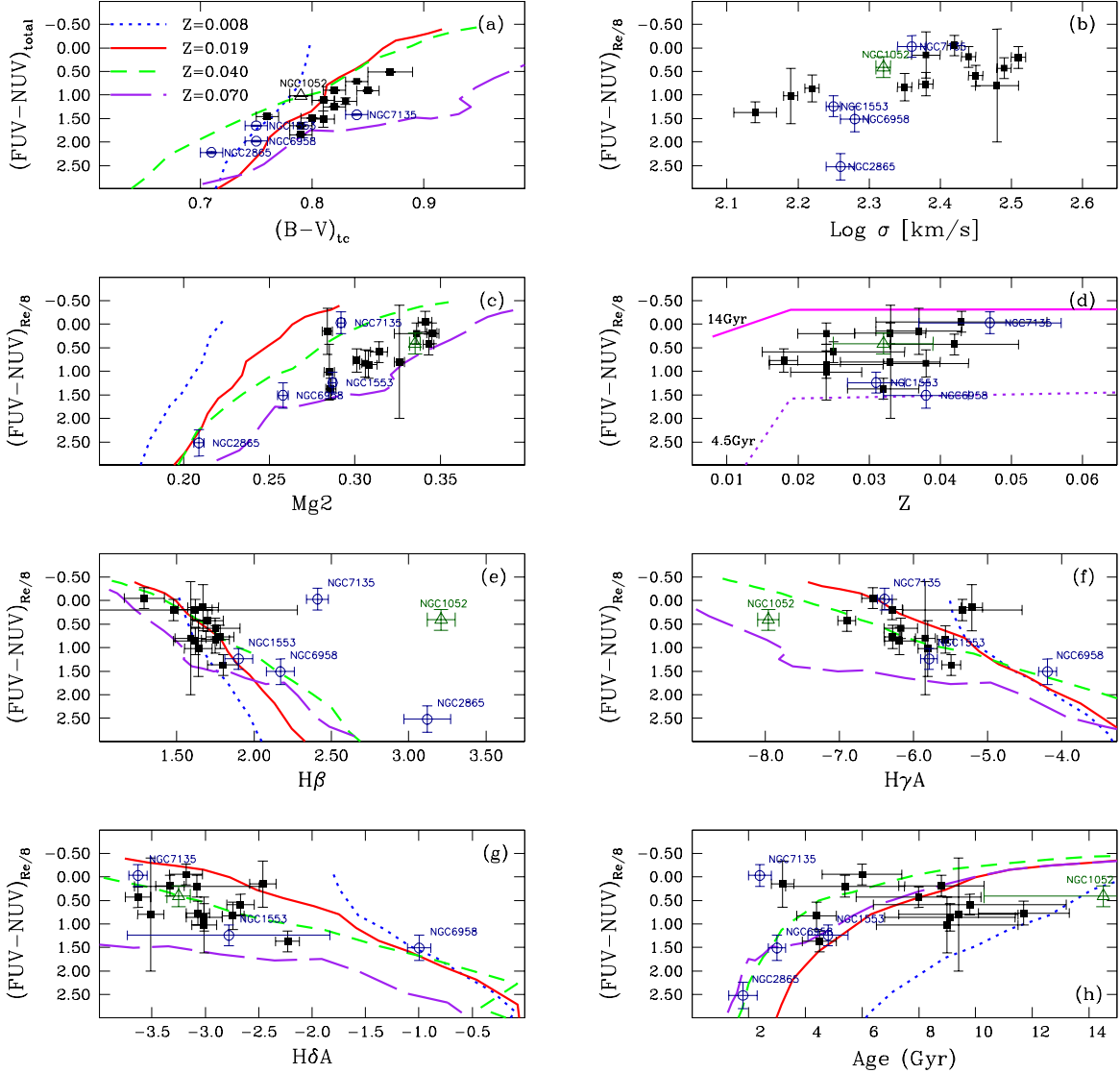}
\caption{{\it (a panel)} {\it GALEX} total (FUV - NUV) colour vs.
corrected total (B-V) colour. (FUV - NUV) colour within an aperture
of r$_{e}$/8 radius vs. central velocity dispersion {\it (b panel)},
Mg2 line-strength index {\it (c panel)}, $Z$ (average galaxy
metallicity) {\it (d panel)}, H$\beta$ {\it (e panel)}, H$\gamma$A
{\it (f panel)}, H$\delta$A {\it (g panel)} line-strength indices
and the the average $Age T_{SF}$ {\it (h panel)} estimated in the
[Age, Z, [$\alpha$Fe] space \citep{Annibali07}. Line-strength
indices have been computed by \citet{Rampazzo05,Annibali06}. Lines
in {\it a, c, d, e, g, h panels} report single stellar population
models of different metallicities (solar metallicity Z=0.019). In
the plane (FUV-NUV)$_{r_e/8}$ vs H$_\beta$ {\it (e panel)}, shell
galaxies, NGC~2865 and NGC~7135, and NGC~1052 appear to be peculiar.
They indeed have quite high H$\beta$ values suggesting the presence
of a young stellar population.  The case of NGC~1052 is probably
connected with the LINER/AGN nature of this elliptical galaxy: its
H$\beta$ line-strength index is objectively difficult to be
corrected for emission line ``infilling" \citep[see
also][]{Annibali07}. NGC~2865 and NGC~7135 could {\it partly} suffer
the problem of a (over)correction of the H$\beta$ line--strength
index (see discussion in $\S~5.4$).  \label{fig10}}
\end{figure*}
%--------------------------end Figure 10-------------------------------------

\subsection{Comparing data with theory}

The comparison of data with theory is made assuming that the complex
stellar mix of a real galaxy can be reduced to a SSP of suitable
metallicity and age. However this approximation has different
implications for the two parameters. While the metallicity
distribution can be ``reasonably'' approximated to the mean value,
the same does not hold for the age, when this derived from
integrated properties \citep[see e.g.][]{Serra07}. In the discussion
below, one has to keep in mind that {\it the age we are measuring
from colours and indices is always biased by the last episode of
star formation}. In other words, it is a mean luminosity weighted
age, in which the most recent star forming episode dominates at
least during the first 2-3 Gyr from its occurrence. This is simply
due to the rather well established law of luminosity fading of
stellar populations, which ultimately mirrors the lifetime and
evolutionary rate of a star as a function of its mass. In
the following we will refer to this age as $T_{SF}$. It does not
necessarily coincide with the real age of a galaxy $T_G$. In
general, if a galaxy underwent an initial star forming episode ever
since followed by quiescence, $T_{SF}\simeq T_G$; if later episodes
of star formation occurred, $T_{SF} < T_G$, the difference getting
larger as the secondary activity gets closer to us in time. An old
galaxy, may look young if secondary star formation activity took
place in the "recent past". We will come back to this in more detail
later on.

The  correlations we have been looking at are displayed in the
panels  (a) through (h) of  Figure~\ref{fig10} in which the galaxies
of Table~\ref{table4} and \ref{table5} and the theoretical
predictions are compared. Among the 18 galaxies under examination,
we label the three of the present study and NGC~1553,  NGC~6958, and
NGC~1052, which deserve particular attention.

According to \citet{Annibali07} NGC~1553 has a luminosity weighted
age $T_{SF}$ of about 4.8$\pm$0.7 Gyr and a metallicity of
Z=0.031$\pm$0.004. The colour (FUV-NUV)$_{r_e/8}$ for this galaxy is
1.24$\pm$0.22. \citet{Rampazzo03} found that ionized gas is present
in the centre of this galaxy. The kinematics shows that the ionized
gas has a nearly regular velocity field suggesting that, if it was
acquired from outside, the gas had time to settle down onto a small
disk.

%--------------------------Figure 11----------------------------------
\begin{figure}
{\includegraphics[width=8.5cm]{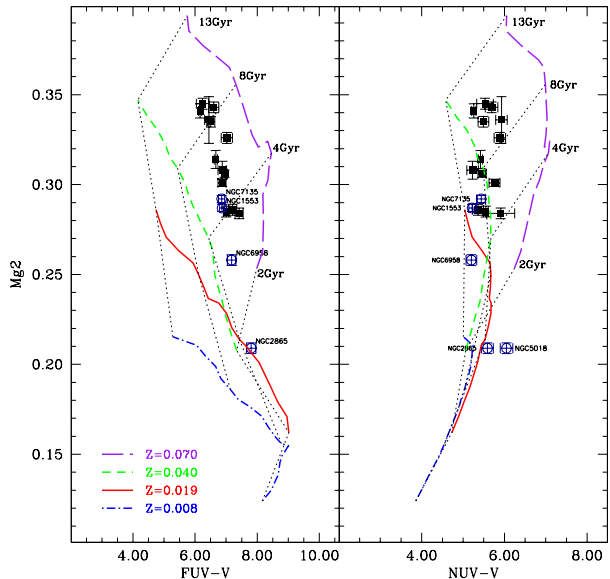}} \caption{ Comparison of
the theoretical relation Mg2 vs. {\it GALEX} UV (FUV on the left and
NUV on the right) - optical colours with our data.
 \label{fig_Mg2_FuvNuv}}
\end{figure}
%--------------------------end Figure11-------------------------------------

NGC~6958 belongs to the \citet{Malin83} list of southern shell
early-type galaxies. \citet{Saraiva99} show that the galaxy does not
show any significant signature of interaction in the elliptically
shaped isophotes. They conclude that, if the galaxy underwent a
merger, the companion galaxy has probably already merged. According
to \citet{Annibali07} the average age $T_{SF}$ of the galaxy in the
central part is about 3.0$\pm$0.3 Gyr and the metallicity is about
two times solar (Z=0.038$\pm$0.006).

Finally, the giant elliptical NGC~1052 is  a well known optical
LINER with an obscured AGN nucleus \citep[see e.g.][]{Terashima02}.

In the plane (FUV-NUV)$_{r_e/8}$ vs (B-V)$_{tc}$ (Figure~\ref{fig10}
panel {\it a}), the galaxies of our sample share the same distribution
of {\it normal} early-type galaxies and shell galaxies are among the
bluest objects \citep[see e.g.][]{Donas06}.  The distributions of
galaxies in the planes (FUV-NUV){$_{r_{e}/8}$} vs. the central
velocity dispersion $\sigma_c$ and Mg2 index (Figure~\ref{fig10} panel
{\it b} and {\it c}, respectively ) are consistent with those reported
for Virgo early-type galaxies in Table~1 of \citet{Boselli05}.

In the plane (FUV-NUV)$_{r_e/8}$ vs H$_\beta$ (panel {\it e} of
Figure~\ref{fig10}), shell galaxies appear to be peculiar. They indeed
have quite high H$\beta$ values suggesting the presence of young
stellar populations. However, only NGC~7135 has also a rather blue
(FUV-NUV)$_{r_e/8}$ colour. The case of NGC~1052 showing also a high
H$\beta$ is probably connected with the LINER/AGN nature of this
elliptical galaxy: its H$\beta$ line-strength index is objectively
difficult to be corrected for emission line ``infilling"
\citep[see also][]{Annibali07}. In this sense,
the shift towards high H$\beta$ of NGC~2865 and NGC~7135
 could  {\it partly} suffer the problem of a (over)correction
of the line-strength index, although, most of the galaxies plotted in
Figure~\ref{fig10} have emission lines in their spectra
\citep[see also][]{Rampazzo05,Annibali06}.

In the planes (FUV-NUV)$_{r_e/8}$ vs H$\gamma$ and H$\delta$, shown in
panels {\it f} and {\it g} of Figure~\ref{fig10}, where corrections
are expected to be lower, shell galaxies more closely follow the trend
and position predicted by the theoretical models.

According to the literature, all shell galaxies in our sample likely
had a secondary, recent burst of star formation in their nuclei. For
NGC~2865 \citet{Longhetti99,Longhetti00} obtained an average
age estimate $T_{SF}$  for this star episode of about 2 Gyr both
from H$\beta$ and the CaII[H+K] indices. A similar age $T_{SF}$
(2.2$\pm$0.4 Gyr) is attributed to NGC~7135 by \citet{Annibali07} as
a consequence of its Balmer line-strength indices and metallicity
higher than 2 times solar (Z=0.047$\pm$0.010). Although these two
galaxies should have the same $T_{SF}$ from line-strength indices
they are located in the age - (FUV-NUV)$_{r_e/8}$) at the two
opposite extremes of the (FUV-NUV)$_{r_e/8}$ values. Notice however
in Figure~\ref{fig6} that NGC~7135 had a radial gradient in
(FUV-NUV) moving from the centre to the outskirts where
(FUV-NUV)$_{r_e/8}$ is approximately 1-1.4. If NGC~7135 hosts a
young stellar population, this is confined in the very centre of the
galaxy. From the data shown in Figure~\ref{fig5} there is no hint
that the burst has occurred also in the shell.

Furthermore, in panels {\it d} and {\it h} of Figure~\ref{fig10} we
show the correlation between (FUV-NUV)$_{r_e/8}$ and the age
$T_{SF}$ and the metallicity $Z$ estimated by \citet{Annibali07}
from the data within r$_{e}$/8 area. Notice that NGC2865 is missing
because \citet{Longhetti00} did not derive the metallicity of this
object. All the galaxies in our sample span large ranges of
metallicity (Z between 0.018 and 0.05) and ages $T_{SF}$(between 2
and 14 Gyr). NGC~7135 is very peculiar because a high metallicity
and a young $Age$ have been estimated for this object: indeed it is
the galaxy with the largest discrepancy from the theoretical
expectation.

Figure~\ref{fig_Mg2_FuvNuv} shows the plane (FUV-V) and (NUV-V) vs Mg2
and the comparison between theory and data. The line--strength index
Mg2 of each galaxy is derived in the r$_{e}$/8 aperture, whereas the
(FUV-V) and (NUV-V) colours refer to the whole galaxy. We are aware
that this makes more difficult the comparison.  Anyway,
\citet{Annibali07} found that, at least within r$_{e}$/2 there are no
gradients in $age$ and/or $\alpha$--enhancements across the galaxies
of their sample, whereas an average negative gradient in metallicity
is firmly detected from the centre to r$_{e}$/2. In any case,
Figure~\ref{fig_Mg2_FuvNuv} shows that a large spread for $T_{SF}$
should exist and that the shell galaxies have the shortest values.

%--------------------------Figure 12----------------------------------
\begin{figure}
{\includegraphics[width=8.5cm]{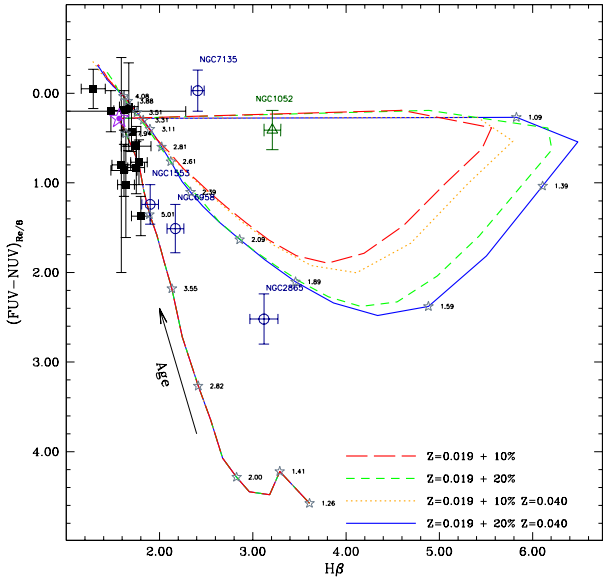}} \caption {{\it GALEX} 
(FUV - NUV) colour  vs. H$\beta$ within an aperture of
r$_{e}$/8 radius. In above plane, the aging path of a galaxy stellar
population is indicated by the arrow. The position of a 10 Gyr old
galaxy with solar chemical composition is indicated by the large star
located in the top left part of the plot. A burst of star formation,
superposed on such an old stellar component, produces the "closed
paths" traced in the plot by dotted, long-dashed, short-dashed and
solid lines according to the percentage of the mass involved in the
burst and of the different metallicity in the newly born stars. The
paths shown consider the percentages of 10\% and 20\% of mass
engaged in star formation with a solar and two times solar
metallicities. The time elapsed from the beginning of the burst is
indicated along the solid line, tracing the path of a burst
involving 20\% of the mass with two times solar chemical
composition. In this context, the time elapsed from the burst of
star formation in  NGC~7135 appears shorter than that of  NGC~2865
and in NGC~1553.
 \label{fig12}}
\end{figure}
%--------------------------end Figure12-------------------------------------

%--------------------------Figure 13----------------------------------
\begin{figure}
{\includegraphics[width=8.5cm]{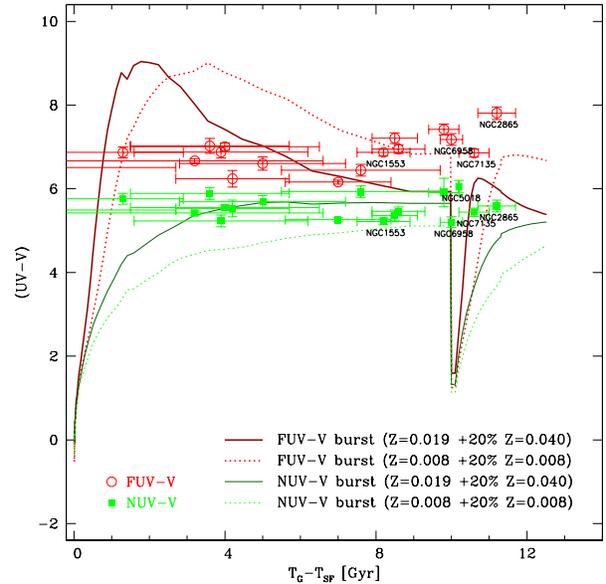} } \caption{{\it
GALEX} UV - optical colours of composite SSPs made by an old
component and a more recent one as a function of the age. 
The age assigned to the real galaxies is $T_G-T_{SF}$=13-$T_{SF}$ (see text). 
The shell galaxies of the present paper are labeled.
The burst occurs at the age of 10 Gyr, as in the previous figure.  
The four cases on display have the same burst intensity (20\%) but differ for 
the metallicity assigned to the and young stars. The colour and symbol codes 
identify the UV-optical colour and the metallicity: the red open circles are for 
(FUV-V) and the green filled squares are for
(NUV-V) of the observed galaxies. The heavy red solid line is for
(FUV-V) of the model Z=0.019+(20\%)0.04, the heavy red dotted line
is for (FUV-V) of the model Z=0.008+(20\%)0.008, the thin green
solid line is for (NUV-V) of the model Z=0.019+(20\%)0.04, and the
thin green dotted line is for (NUV-V) of the model Z=0.008+(20\%)0.008. 
Note the different recovery time and amplitude of the post-burst colour 
variation at changing metallicity.  
\label{fig13}}
\end{figure}
%--------------------------end Figure13-------------------------------------

\subsection{Simulating bursts of star formation}
%\bf{
\textit{Composite models with bursts}. The above comparison of data
with theory has repeatedly strengthened the hint that at least part
of the scatter in the various diagnostic planes could be due to
recent episodes of star formation (see Figure~\ref{fig_Mg2_FuvNuv}
and the panels $d$, $e$, and $h$ of Figure~\ref{fig10}). Not all
diagnostic planes have the same power; however, the plane (FUV-NUV)
vs. H$\beta$ (panel $e$ of Figure~\ref{fig10}) deserves particular
attention because of the peculiar position of NGC 7135, NGC 1052 and
NGC 2865.

With the aid of the population synthesis technique we present
simulations of composite stellar populations in which a recent burst
of arbitrary intensity (mass percentage engaged in star formation) and
arbitrary age is superposed to an old stellar component with typical
age of 10 Gyr and solar-like chemical composition.  The simulations
can also account for different metallicity between the old and newly
born stars. All details of the technique in use can be found in
\citet{ChiosiCarraro2002} and \citet{Tantalo04b}.

Results of these simulations are shown in
Figure~\ref{fig12} in which is plotted the (FUV -
NUV) colour vs. H$\beta$ within an aperture of r$_{e}$/8 radius. The
lines describe the Age path induced by a recent burst of star
superposed to a 10 Gyr old galaxy stellar population. The recovery
time (i.e. the time elapsed from the beginning of the burst to the
recover of the pre-burst situation, i.e. colours, line indices, etc)
of these simulations is about 2 Gyr or so. The analyses made by
\citet{ChiosiCarraro2002} and \citet{Tantalo04b} show that the
recovery time increases with the percentage of mass engaged in star
formation during the burst. Percentages above 20 \% are not likely
because the recovery time would be a significant fraction of the
Hubble time and accordingly numerous galaxies could be caught in the
bursting mode. See also \citet{Longhetti00} for similar simulations
and conclusions. In this context, the burst in NGC~7135 (and
NGC~1052, see anyway the above notes about the nature of this
galaxy) seems more recent than NGC 2865, NGC~6958 and NGC~1553.
These latter are becoming {\it normal} ellipticals. Therefore, the
anomalous position of the three galaxies in question can be
explained.

\textit{Anomalous position of a few objects in diagnostic planes}.
It is also worth noting that the galaxies with anomalous position in
Figure~\ref{fig12} also have anomalous positions in
panel $h$ of Figure \ref{fig10}. The situation is best illustrated
in Figure~\ref{fig13} (analog of panel $h$) which
correlates the (FUV-V) and (NUV-V) to the age. The models show the
colour evolution with a recent burst of star formation. The burst
occurs at the age of 10 Gyr. The intensity of burst is 20\% and two
possible combinations metallicity for the old and young components
are shown. Of course, other values of the age at which the burst
occurs, its intensity, and the chemical composition could be
adopted. The age assigned to the observed galaxies is $T_G-T_{SF}$,
where for $T_G$ we adopt the canonical age of 13 Gyr. In other words
all galaxies are supposed to initiate their star formation history
13 Gyr ago. Therefore the plotted age is 13-$ T_{SF}$ Gyr. The aim
of this plot is to compare the observed properties of the galaxies
in our sample with a typical star formation history of a galaxy as
suggested by the analysis carried out in the previous section. The
choice of age at which the burst of star formation occurred is
suggested by the properties, i.e. (FUV-NUV) colours and $H\beta$
indices of the shell galaxies. Attempts to reproduce other galaxies
in the sample would require different assumptions for the burst age,
but this is beyond the aims of this study.

 There are
at least five results out of this comparison:

(i) In general the strongest variations of the colours are in coincidence 
of or soon after the star forming episodes (the initial one, 
$T_{SF}\simeq T_G$, and the burst, $T_G-T_{SF}$=10).

(ii) The colour (FUV-NUV) is sensitive both  to metallicity
and age over a long fraction of a galaxy life (see Fig~\ref{fig10} panel
$h$);

(iii) The colour (NUV-V)  (Fig~\ref{fig13}) soon becomes nearly  age 
independent, say 2 -- 3 Gyr past the star forming activity both
after the initial episode and the burst.  This time interval
increases with the fractionary mass engaged in the star forming
episode: it is rather long in the initial episode (up to 4 Gyr) and
only 1-2 Gyr after a burst of moderate intensity. It runs nearly
flat during the long time interval between the two star forming
episodes (in this simulation) see also \citet{Yi05} and \citet{Kaviraj06}. 
It keeps, however, some dependence on the metallicity,  shifting to
higher values at increasing Z;

(iv) The colour (FUV-V) (Fig~\ref{fig13}) keeps a good dependence 
on both age and metallicity all over the galaxy life;

(v) The anomalous location of the few shell galaxies under
consideration correspond to the post-burst variations of the colour.
By finely tuning the burst age, intensity and metallicity and the
chemical composition of the old component we could easily match
individual galaxies. This is beyond our present aims.

\begin{figure}
{\includegraphics[width=8.5cm,height=12.5cm]{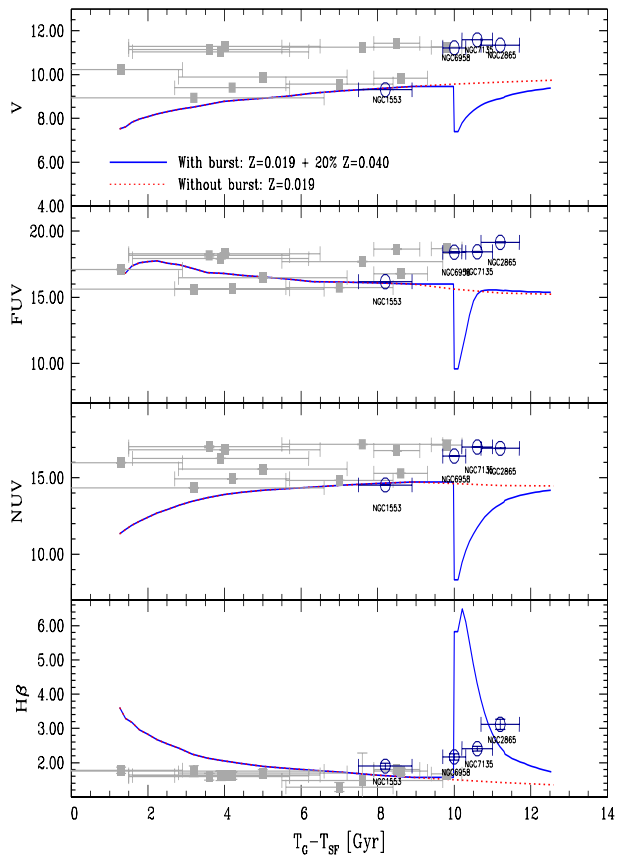}}
\caption{{\it GALEX} FUV, NUV, V  magnitudes, and $H\beta$
index as a function of the age in a model galaxy evolved in quiescence 
(dotted lines). The theoretical FUV, NUV, and V magnitudes have been 
rescaled to the faint end boundary of the data (see text). The
 thick symbols show the same quantities but for the model galaxy in
 which a recent burst of star formation is added. The burst occurs at
 10 Gyr. In this simulation the original metal content is Z=0.019, the
 mass fraction in the burst is 20\% and the metallicity of the newly
 born stars is Z=0.04.  The age plotted for the observed galaxies is
 $T_G -T_{SF}$ defined as in the previous plot.
 \label{fig14} }
\end{figure}

The temporal behaviour of the (FUV-V) and (NUV-V) colours as
function of the age both in the pre- and post burst regimes can be
easily understood looking at the separate variation of the AB
magnitudes FUV, NUV and V of our model galaxies with and without
bursts of activity as a function of the age that are shown in the
three top panels of Figure~\ref{fig14} which does not require any
additional explanation. The only thing to mention is that
the magnitudes of the model galaxies (in reality composed SSPs) have 
been vertically shifted to make them nearly coincident with the
observed ones. The vertical shift simply accounts for the difference
between the luminous mass of the galaxies and the much smaller mass 
of the SSPs. The shift (the same for all pass-bands) is such that
theoretical models and real data coincide at the faint end boundary
of the data. In principle, an estimate of the mass could be derived
from the velocity dispersion. However, as this would require some
modelling of the relative mass in Dark and Baryonic Matter, we
prefer to simply scale the magnitudes of the composite SSPs by a
suitable amount. This is fully adequate to our purposes. Also in
this case, the age adopted for the observed galaxies is
$T_G-T_{SF}$=13-$ T_{SF}$ Gyr. The agreement is fairly good. The
shell galaxies on display can be associated to the post-burst period
as the detection probability is higher thank to the slower fading
rate. Perhaps they suggest some individual fine tuning of the burst
age, which is, however, beyond our aims.

Finally in the bottom panel of Figure~\ref{fig14} we show the temporal
evolution of $H\beta$ of the model galaxies with the observations. No
vertical shift is now required (because the line strength indices are
by definition mass independent).  The age assigned to each real galaxy 
is the same as before. The agreement is remarkably good and the same 
remarks on the fine tuning of the burst age to fully match the shell 
galaxies could be made.

\begin{figure}
{\includegraphics[width=8.5cm]{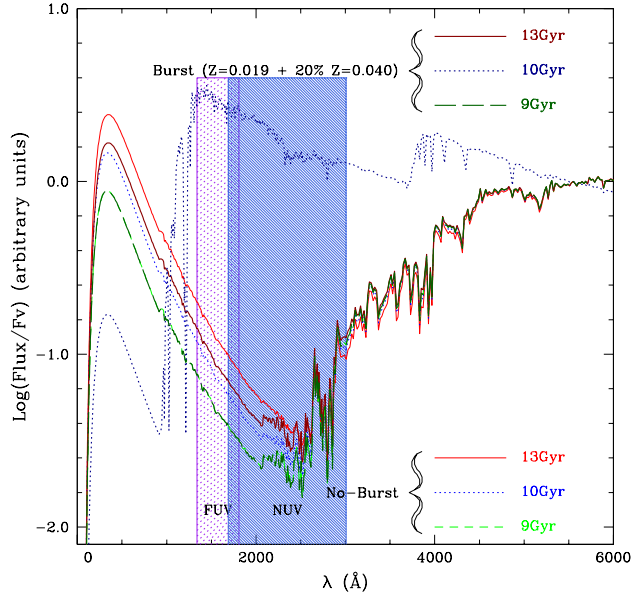}} \caption{
Spectral energy distributions as a function of the age (9,
10, 13 Gyr) for quiescent model galaxies (thin lines) and galaxies
with a recent burst of star formation (heavy lines). The burst
occurs at 10 Gyr, and fades afterward. In this simulation the
original metal content is Z=0.019, the mass fraction in the burst is
20\% and the metallicity of the newly born stars is Z=0.04. The
wavelength intervals for the {\it GALEX} FUV and NUV pass-bands 
are indicated by the two shaded areas. 
\label{fig15} }
\end{figure}

\textit{Which is the source of the FUV and NUV fluxes?} The
occurrence of a "rejuvenation" process in an early type galaxy by a
recent burst of stars raises the issue of disentangling between the
different sources (old and/or young stars) producing the FUV and NUV
fluxes. Perhaps the best way of addressing this issue is to look at
the spectral energy distribution of our model galaxies and its
variation with time. This is shown in Figure~\ref{fig15}. A
quiescent galaxy (of sufficiently high mean metallicity), during the
time interval from 9 to 13 Gyr, changes its spectrum as shown by the
thin lines in Figure~\ref{fig15}: at wavelengths longer than
about 2500 \AA\ little action occurs, whereas short-ward of it the
UV flux in the up-rising branch gets stronger. The slope of the
up-rising branch below 2000 \AA\ is nearly constant. Most likely,
the source of this flux are the old hot-HB, and/or post AGB, and the
high-metallicity, AGB-manqu\'e stars \citep[see][ for more
details]{Bressan94}. The NUV region contains both flux from MS
turn-off stars of the evolved population \citep[see
e.g.][]{Dorman03}, perhaps independently of their extension across a
galaxy, and flux from the UV-upturn. In contrast, the FUV emission
is due only to the evolved stars. In presence of a burst, the
situation is more complicated. In the example shown in
Figure~\ref{fig15}, we start with the spectrum at 9 Gyr exactly
the same as the previous case, we jump to the spectrum dominated by
the young component (heavy dotted line) at the time of the burst
occurrence (10 Gyr), and then we nearly fall back to the previous
case at the age of 13 Gyr. But for minor difference in the UV rising
branch (short-ward of 2000 \AA), the spectrum is the same as in the
quiescent case (recovery of the old situation). The trace of the
burst of stellar activity is limited to the minor difference in the
flux of the rising branch. It is obvious that changing the age and
the intensity of the burst many intermediate situations can be found
in which the FUV and the NUV are more dominated by the recent burst of
stellar activity. All this is of course mirrored by the colours and
line strength indices in the typical diagnostic planes. Owing to the
rapid fading of the burst, the probability of catching a galaxy
during the maximum stellar activity so that the FUV and the NUV
pass-bands are filled by radiation from the newly born stars is low
compared to the probability of catching a galaxy either in the pre-
or post-burst regime. This type of analysis holds its validity only
in statistical sense. To conclude there is no unique answer
to the question "which is the source of the FUV and NUV fluxes?"
unless more information is provided.

%--------------------------Figure 14----------------------------------
%--------------------------end Figure14-------------------------------------

%--------------------------Figure 15----------------------------------
%--------------------------end Figure15-------------------------------------

\section{Summary and conclusions}

We have presented the data for three early-type nearby galaxies (NGC
2685, NGC 5018 and NGC 7135) observed in the {\it GALEX} FUV and NUV
pass-bands. These galaxies are characterized by prominent shell
systems. The NUV emission of NGC~7135 and NGC~2685 mirrors the optical 
appearance and features. This implies that the UV emission comes from the
 same kind of stellar populations. The FUV images are instead more 
 concentrated toward the nucleus in the case of NGC~2865 and/or in the
case of NGC 7135 other particular regions of the galaxy thus
suggesting that different types of hot stars contribute to the
flux. In the case of NGC~5018, for which only the NUV image is
available, it is not possible to reveal the shell structure visible in
the optical, whereas the presence of a complex dust-lane system is
seen.

All the galaxies show evidence of dust features in their centre. This
confirms the trend found by \citet{Sikkema07} who suggest that the
dust detection rate in the shell ellipticals is significantly higher
than in a ``normal" elliptical sample. Furthermore the irregular shape
of shell galaxies suggests that external influences and/or acquisition
of material are the cause of the dust features.

The three galaxies seem to belong to loose, poor groups rich of cold
gas whose presence is traced by the H\,{\sevensize I} emission.  We
suggest that these objects typically occur in {\it evolving groups}
similar to the Arp~227 group dominated by the shell galaxy NGC~474
\citep{Rampazzo06}. No galaxy of our sample shows the presence of cold
gas in its innermost regions. In the case NGC~2865
\citet{Schiminovich95} find that the outer cold gas is kinematically
associated to the stellar body. This strongly suggests that whatever
the phenomenon giving rise to the shells might be (the agreement
between the gas and stellar kinematics somewhat rejects the merging
hypothesis) it is able to convert the gas in the central regions into
stars, thus ``rejuvenating" the galaxy in agreement with what inferred
from the line--strength indices.

The analysis of the colour evolution of single and composite
stellar populations simulating past and/or recent star formation
shows that  (i) the colour (FUV-NUV) has good sensibility to
metallicity and age over a long fraction of a galaxy life;(ii) The
colour (NUV-V)  soon becomes nearly  age independent, say 2 -- 3 Gyr 
past the star forming activity both after the initial episode and
the burst.  This time interval increases with the fractionary mass
engaged in the star forming episode. So it is rather long in the
initial episode (up to 4 Gyr) and only 1-2 Gyr after a burst of
moderate intensity. It runs nearly flat during the long time
interval between the two star forming episodes (in this simulation). 
It keeps, however, some dependence on the metallicity shifting to higher 
values at increasing Z; (iv) finally, the colour (FUV-V) keeps a good 
dependence on both age and metallicity all over the galaxy life. 
In the typical case of secondary activity of moderate intensity, all
the colours but for (FUV-NUV) and (FUV-V),  become nearly  age
insensitive when 1-2 Gyr have elapsed from the last star forming
episode \citep{Jeong07}. 

Most likely, the NUV and FUV fluxes of ``normal'' early-type galaxies
have different origins. Independently of its extension across a galaxy, the NUV flux
is partially due to the MS turnoff stars of the evolved population
\citep[see e.g.][]{Dorman03} and partially to more evolved, exotic stars.
In contrast, a strong FUV emission is likely due to the presence of
one (or more) hot, plausibly high-metallicity, stellar components
giving origin to the well-known phenomenon of the UV-upturn
(e.g. hot-HB and/or post-AGB and AGB manqu\'e stars). Further
complication comes from the analysis of the optical indices, which
suggest that the three shell galaxies in our study had a recent (2-3
Gyr old) burst of stellar activity, possibly as a consequence of the
interaction/accretion episode that triggered the shell formation. In
this context, it comes naturally that the stellar populations should
span a certain range of ages and also that young stars could be
present in different proportions in both the NUV and FUV fluxes. 
All this makes the interpretation of the data more difficult.

Our combined analysis of the (FUV-NUV) and (UV-V) colours and
line--strength indices shows that even simple models such as SSPs
mimicking the  passive evolution of a galaxy can account for  most
of the gross features  of shell galaxies as well as those of normal
early-type galaxies. Among the shell galaxies in the sample, the
nucleus of NGC~7135 shows the more peculiar behavior when compared
with theoretical expectation of passive evolution.  
Considering composite stellar population models with a recent burst
of star formation, we show that the position of the NGC~7135 nucleus
in the (FUV-NUV)-H$\beta$ plane could be explained in term of a
recent rejuvenation episode. NGC~2865 and the other few shell
galaxies, which have a nearly ``normal'' position in the above
plane, could also fit the same rejuvenation framework.

Recently \citet{vanDokkum05} suggests that
{\it dry} mergers, i.e. nearly dissipationless, or gasless mergers,
at low redshift are responsible for much of the local bright field
elliptical galaxy population. \citet{Donovan07} demonstrated that
the morphological and photometric criteria adopted by
\citet{vanDokkum05} to identify {\it dry} merger candidate could
include early-type galaxies with H\,{\sevensize I} in or around them
i.e. possible {\it wet} ellipticals. The sample they considered is
taken from the H\,{\sevensize I} Rogues Gallery \citep{Hibbard01}
and include NGC~2865, NGC~5018 and NGC~7135 which all fit the
criteria for being {\it dry} mergers candidate. Our results suggest
that the merger that has induced a secular evolution in NGC~7135
(and possibly in NGC 2865) should not have been so {\it dry}.

\section*{Acknowledgments}
This research has been partially founded by ASI-INAF contract
I/023/05/0. {\it GALEX} is a NASA Small Explorer, launched in April
2003. {\it GALEX} is operated for NASA by California Institute of
Technology under NASA contract  NAS-98034. {\it GALEX} Guest
Investigator program GALEXGI04-0030-0059. This research has made
use of the NASA/IPAC Extragalactic Database (NED) which is operated
by the Jet Propulsion Laboratory, California Institute of
Technology, under contract with the National Aeronautics and Space
Administration. The Digitized Sky Survey (DSS) was produced at the
Space Telescope Science Institute under U.S. Government grant NAG
W-2166. The images of these surveys are based on photographic data
obtained using the Oschin Schmidt Telescope at the Palomar
Observatory and the UK Schmidt Telescope. The plates were processed
into the present compressed digital form with the permission of
these institutions.

\label{lastpage}

\end{document}